\documentclass[reprint,groupedaddress,eqsecnum,amsfonts,amsmath,amssymb,aps,prd]{revtex4}
\usepackage{graphicx}
\usepackage{latexsym}
\usepackage{amssymb}
\usepackage{mathrsfs}
\usepackage{amsmath}
\usepackage{amsthm}
\usepackage{upgreek}
\usepackage{color}
\usepackage{dcolumn}
\usepackage{bm}

\definecolor{dyellow}{rgb}{1.,0.8,.0}
\definecolor{myblue}{rgb}{.1,.1,.7}
\definecolor{dcyan}{rgb}{.0,.6,.6}
\definecolor{dmagenta}{rgb}{0.6,0.0,0.6}
\definecolor{brown}{rgb}{0.6,0.2,0.}
\definecolor{darkblue}{rgb}{.0,.0,0.5}
\definecolor{darkred}{rgb}{0.75,0.0,0.0}
\definecolor{orange}{rgb}{1.,.6,.0}
\definecolor{dorange}{rgb}{0.8,.4,.0}
\definecolor{darkgreen}{rgb}{0.0,0.6,0.0}
\definecolor{purple}{rgb}{.4,.0,.4}
\definecolor{grey}{rgb}{0.5,0.5,0.5}

\begin{document}
\hyphenpenalty=1000
\preprint{APS/123-QED}
\title{Motion of extended fluid bodies in the Newtonian limit of $f(R)$ gravity}

\newcommand*{\PKU}{Guangxi Key Laboratory for Relativistic Astrophysics, School of Physical Science and Technology, Guangxi University, Nanning 530004, China}\affiliation{\PKU}
\newcommand*{\INFN}{Guangxi Center for Mathematical Research, Guangxi University, Nanning 530004,  China}\affiliation{\INFN}
\newcommand*{\CICQM}{Academy of Mathematics and Systems Science, Chinese Academy of Sciences, Beijing 100190, China}\affiliation{\CICQM}
\newcommand*{\CHEP}{School of Mathematical Sciences, University of Chinese Academy of Sciences, Beijing 100049, China}\affiliation{\CHEP}

\author{Bofeng Wu}\email{wubofeng@gxu.edu.cn}\affiliation{\PKU}
\author{Xiao Zhang}\email{xzhang@amss.ac.cn}\affiliation{\INFN,\CICQM,\CHEP}

\begin{abstract}
In the Newtonian limit of $f(R)$ gravity, for an isolated self-gravitating system consisting of $N$ extended fluid bodies, the inter-body dynamics are studied by applying the symmetric and trace-free formalism in terms of irreducible Cartesian tensors. The multipole expansion of each body's center-of-mass acceleration is derived, and the expansion comprises the Coulomb-type part and the Yukawa-type part, where the former, identical to that in General Relativity, is encoded by the products of the mass multipole moments of the body with those of other bodies, and the latter, as the modification introduced by $f(R)$ gravity, is encoded by the products of the scalar multipole moments of the body with those of other bodies. Due to the finite range of the massive scalar mode, the scalar multipole terms in the Yukawa part do not decay systematically with increasing order. As an essential component of the system's orbital dynamics, the multipole expansion for the total gravitational potential energy is provided, and the expression for the total conserved energy in terms of the mass and scalar multipole moments of the bodies is offered. To investigate the system's spin dynamics, the equation of motion for each body's spin angular momentum is further deduced and presented in the form of multipole expansion. These findings constitute the main content of the coarse-grained description of inter-body dynamics for the system within the framework of the Newtonian limit of $f(R)$ gravity. As a by-product, for a two-body system, the effective one-body equation governing the relative motion between the two bodies and the total energy of this system are achieved.
\end{abstract}
\maketitle
\section{Introduction}
\label{Sec:first}
\textbf{Theoretical Background and Literature Review\ ---} Although General Relativity (GR) has achieved astonishing accomplishments~\cite{Eric2014,Clifford2018,TheLIGOScientific:2016agk,TheEHTcientific:2019agk},
it still faces many challenges when it is used to interpret the observed data at astrophysical and cosmic scales~\cite{cosmicacceleration}.
It is well-known that one approach to addressing these challenges is to construct alternative theories of gravity, and these theories are based on generalizations of GR~\cite{Capozziello:2011et,Clifton:2011jh,DNojiri:2017ncd,Johnson2022}. In this paper, we will focus our attention on $f(R)$ gravity~\cite{Sotiriou2010,DeFelice:2010aj,Nojiri:2010wj,Scali2024}, and as a typical relativistic gravity theory, it replaces the Ricci scalar $R$ in the Einstein-Hilbert action with a general function $f$ of $R$.
Compared with GR, $f(R)$ gravity can not only improve renormalization properties~\cite{Stelle:1976gc} but also naturally lead to a period of accelerating expansion early in the universe's history~\cite{Starobinsky:1980te}, so it could be considered as an attractive candidate for the alternatives to GR~\cite{Clifton:2011jh}. To test $f(R)$ gravity with astrophysical observations, it is essential to study the motions of extended bodies in such models. Given that the Yukawa-type modification to the Newtonian potential introduced by such models~\cite{Stabile:2010zk,Naf:2010zy,DeMartino:2018yqf,DeLaurentis:2018ahr} renders point-mass approximations invalid, in celestial mechanical systems involving extended sources, the motion of an object is heavily influenced by the size and shape of the gravitational source.

The theoretical treatment of the motions of extended bodies in gravitational fields has a rich history in metric theories of gravity~\cite{Kopeikin2011,Soffel2019,Damourhistory,Blanchethistory}. These developments include the systematic formulations of relativistic celestial mechanics and reference systems by Kopeikin, Efroimsky, and Kaplan~\cite{Kopeikin2011}, and by Soffel and Han~\cite{Soffel2019}, as well as the multipolar post-Newtonian frameworks developed by Damour and collaborators~\cite{Damourhistory} and by Blanchet and collaborators~\cite{Blanchethistory}.  Collectively, these all provide the foundation for the classical description of extended-body dynamics in GR and scalar-tensor gravity.  Complementing the classical description, recent advances in the Effective Field Theory (EFT) of gravity offer a powerful and systematic alternative for analyzing the dynamics of extended bodies. Pioneered by Goldberger, Rothstein, and Ross~\cite{Goldberger2006,Ross2012,Goldberger2010}, this approach utilizes field-theoretic methods to treat extended bodies as worldline point particles endowed with multipole moments. Within the EFT framework, the finite-size effects and spin of extended bodies are systematically encoded into worldline operators. Notably, this formalism not only reproduces the classical results but also provides a more efficient computational scheme for complex effects such as gravitational radiation reaction and finite-size corrections.

\textbf{Physical Motivation and Model Assumptions\ ---} The gravitational potential in $f(R)$ gravity consists of a Newtonian part and a Yukawa correction, representing the massless and massive scalar fields~\cite{Martino2018,Toniato2018,Capozziello2010}. While the multipole expansion of a massless scalar field is straightforward, the massive scalar field introduces significant complications. Thus, a systematic multipole analysis of extended-body dynamics in a self-gravitating system under $f(R)$ gravity has not been developed. EFT has been applied to scalar-tensor gravity to study extended-body motion and gravitational radiation~\cite{Kuntz2019,Kopeikin2019}. However, those studies involve only massless fields and do not provide the required multipole analysis for $f(R)$ gravity. Refs.~\cite{Wu:2021uws,Wu:2023qjp} give the metric for an extended source's external field up to $1/c^3$ order in $f(R)$ gravity in the form of multipole expansion. From this, the potential's multipole expansion can be read off. It includes GR-like and modified terms, determined by mass and scalar multipole moments. Since all scalar multipole moments depend on the mass distribution, unlike in GR, a spherical body's external field cannot be generated by a point source at its center~\cite{Castel-Branco:2014exa}. Thus, in $f(R)$ gravity, the source's size and shape crucially affect near-field phenomena. Here, building on Refs.~\cite{Wu:2021uws,Wu:2023qjp}, we aim to perform a complete multipole analysis of extended-body dynamics in the Newtonian limit of $f(R)$ gravity.

In this paper, for an isolated self-gravitating system of $N$ extended fluid bodies, we will study the inter-body dynamics under the Newtonian limit of $f(R)$ gravity using the classical method. While the EFT framework offers a powerful tool for higher-order PN corrections and radiation reactions, the classical approach is sufficient for our focus on the Newtonian limit. The self-gravitating system is assumed to satisfy two assumptions: 1) Each body is composed of a perfect fluid, and maintains approximate hydrostatic equilibrium; 2) No matter is being ejected or accreted by any body, and their typical sizes are far less than the typical separation between them.
Although these two assumptions are applicable to a wide range of  self-gravitating systems, such as binary systems of main-sequence stars, they are not intrinsic requirements from the perspective of general extended-body dynamics. Rather, they are simplifying assumptions adopted within the present fluid-based framework, not prerequisites for the general theory of extended-body motion. To clearly delineate the range of applicability of the results obtained in this paper, the validity and scope of the above assumptions  are examined in detail below.

The first assumption provides a tractable framework that decouples the intra-body dynamics from the inter-body dynamics, enabling
a coarse-grained description of each body in terms of total mass, center-of-mass position, spin angular momentum, and multipole moments. However, this assumption inherently excludes more complex internal rheology or violent dynamical evolution, so the framework does not apply to systems undergoing tidal disruption or strong pulsations. The second assumption ensures mass conservation for each individual body, maintains well-defined and time-independent integration domains, and provides the correct condition for the multipole expansion of the integral kernels in the center-of-mass frame. Therefore, this assumption renders our framework inapplicable to close binaries with active mass transfer, or to systems approaching collision or merger. It is worth emphasizing, however, that while the perfect-fluid assumption is adopted for the derivation, the Newtonian-order gravitational field depends only on the macroscopic mass density distribution and is independent of the specific equation of state or internal composition. Thus, the resulting inter-body dynamics of this paper are broadly applicable to extended bodies composed of arbitrary matter remaining in mechanical equilibrium, whether fluid or solid, such as Earth-like planets.

\textbf{Translational Motion and Specific Problem\ ---} To understand the orbital dynamics of the system, it is essential to derive the equations of motion for all the bodies, which necessitates specifying their center-of-mass accelerations. Since $f(R)$ gravity is considered only at the Newtonian order, the definition of an object's center-of-mass acceleration is identical to that in classical mechanics. In general, each body's center-of-mass acceleration is obtained by inserting Euler's equation of hydrodynamics in a gravitational field and the gravitational potential produced by all the bodies in the system into its definition. However, performing a multipole analysis on the center-of-mass acceleration of each body by applying the symmetric and trace-free (STF) formalism  in terms of irreducible Cartesian tensors~\cite{Thorne:1980ru,Blanchet:1985sp,Blanchet:1989ki,Damour:1990gj} presents a challenging and critical issue. Specifically, the primary technical difficulty lies in performing a multipole analysis on the integral kernel and its spatial derivative for the screened Poisson operator in the center-of-mass frame of each body within the system. In this paper, we resolve this challenge by utilizing spherical modified Bessel functions and the STF formalism to successfully decouple the position vector of each body with respect to its center of mass from the relative center-of-mass separation vectors, and present the corresponding results, namely closed-form multipole expansions for the Yukawa-type kernel and its spatial derivative given in Eqs.~(\ref{equ3.48}) and (\ref{equ3.52}). Consequently, the multipole expansion for the center-of-mass acceleration of each body in the system is also derived (which, as discussed below, exhibits a dual structure originating from the Coulomb and Yukawa kernels, as well as non-decaying scalar multipole terms that are rooted in the Yukawa kernel).

The expansion contains the Coulomb-type part and the Yukawa-type part, which are, respectively, contributed by the GR-like terms and the modified terms in the gravitational potential. The Coulomb-type part is encoded by the products of the mass multipole moments of the body with those of other bodies and is exactly the same as its counterpart in GR. The Yukawa-type part is encoded by the products of the scalar multipole moments of the body with those of other bodies and represents the modification to the Coulomb-type part introduced by $f(R)$ gravity. Given that the multipole expansions for the center-of-mass accelerations of all the bodies form a complete set of equations of motion for the orbital dynamics of the system once the mass and scalar multipole moments of the bodies are specified, they provide the foundation for understanding the orbital motion of the system within the framework of $f(R)$ gravity.

If the system is composed of ideal spheres, the multipole expansions for the center-of-mass accelerations of the bodies only retain the monopole terms. This result shows that in $f(R)$ gravity, due to the presence of the scalar monopole term, the center-of-mass acceleration of a spherical body differs essentially from that of a point source. This stands in contrast to the case of a massless scalar field,
where the equivalence between a spherical body and a point source, a standard result in GR, is not a generally valid statement in scalar-tensor theories. Rather, the notion of sphericity itself depends on the specific definition of multipole moments adopted in such theories. This distinction highlights a key distinguishing feature of $f(R)$ gravity: regardless of definitional subtleties, it is the finite range of the massive scalar field that intrinsically makes the body size relevant even for spherically symmetric configurations. For non-spherical bodies, the multipole expansion for each body's center-of-mass acceleration reveals that its orbital motion is influenced by several factors: the deformation-induced gravitational potentials generated by other bodies, its own non-spherical mass distribution coupled to the monopole potentials caused by other bodies, and the interactions between its own higher-order multipole moments and those of other bodies. Of these factors, the deformations and non-spherical mass distributions are characterized by higher-order mass and scalar multipole moments, so this suggests that both the size and shape of the bodies can significantly impact orbital motion under $f(R)$ gravity. These multipole couplings also appear in GR and massless scalar-tensor theories. However, in GR, the mass dipole moment vanishes and higher-order mass multipole terms decay with order. While non-zero scalar dipole moments also appear in massless scalar-tensor theories, the truly distinctive feature in $f(R)$ gravity is that scalar multipole terms do not follow the standard hierarchical decay with order (due to the Yukawa kernel), which breaks the conventional power-counting and truncation expectations.

\textbf{Conserved Quantities and System Energies\ ---} For the system, we next demonstrate the conservation of the total momentum, energy, and angular momentum in the Newtonian limit of $f(R)$ gravity, and present their expressions in the coarse-grained description of the system. When examining the system's total energy, a multipole analysis of the total gravitational potential energy is necessary, because it depends on the gravitational potential contributed by all the bodies. In this work, we derive the multipole expansion for the system's total gravitational potential energy. It is shown that the potential energy comprises two types of interaction energies between different bodies: one involving the couplings of the mass multipole moments and the other involving the couplings of the scalar multipole moments. With the expansion of the gravitational potential energy, the expression for the system's total energy is also obtained. In this expression, the monopole-monopole potential energy indicates that the total energy of the system composed of ideal spheres is distinct from that of the system made up of point sources. This implies that unlike in GR, the sizes of the spherical bodies make an important contribution to the system's total energy in $f(R)$ gravity. In addition to the monopole-monopole potential energy, the energy of the system is also contributed by all the monopole-multipole and multipole-multipole potential energies between different bodies. From these results, it is clear that the monopole-dipole and dipole-dipole potential energies are essential for the system's energy under $f(R)$ gravity, whereas they do not play a role in GR as the mass dipole moments of bodies are zero. It should be noted that monopole-dipole and dipole-dipole potential energies also appear in some massless scalar-tensor theories~\cite{Kopeikin2019}. However, the scalar field in these theories is long-ranged, in contrast to that in $f(R)$ gravity, where the scalar field is short-ranged due to the Yukawa suppression. This finite range leads to the non-hierarchical behavior of higher-order scalar multipole terms, which is a distinctive feature of the massive scalar field.

\textbf{Spin Evolution and Weighted Multipole Moments\ ---} To further investigate the system's spin dynamics, the time derivative of each body's spin angular momentum is derived and presented in the multipole expansion form under $f(R)$ gravity at the Newtonian order. This formula describes the time evolution of each body's spin angular momentum and provides
the equation of motion governing its spin behavior. Therefore, the equations of motion for the spin angular momenta of all the bodies serve as the starting point for understanding the system's spin motion in $f(R)$ gravity.
Similar to the acceleration case, the multipole expansion for the time derivative of each body's spin angular momentum also consists of the Coulomb-type and Yukawa-type parts, and they also arise from the interactions between the multipole moments of the body and those of other bodies.
From this expansion, it is important to note that the body's monopole moment does not influence its spin's motion, which is consistent with the inherent properties of the spin angular momentum and presents an obvious distinction between spin and translational acceleration. Apart from this, another key distinction between them lies in the fact that compared to the conventional definition employed in the multipole expansion for the center-of-mass acceleration, the definition of the scalar multipole moments involved in the multipole expansion for the time derivative of the spin angular momentum needs a slight extension to incorporate the concept of weight  (cf.~Eq.~(\ref{equ4.53})), which actually reflects the intrinsic nature of the spin angular momentum.

\textbf{Two-Body Systems and Effective-One-Body Reduction\ ---} The application of this paper's previous findings to studying the orbital dynamics of a two-body system is pivotal for testing $f(R)$ gravity. Just as in celestial mechanics, the orbital dynamics of a two-body system are often explored using the effective one-body approach. To facilitate future applications, in this paper, we derive the effective one-body equation governing the relative motion between the two bodies and the expression for their total energy. With these results, the unperturbed orbits of a two-body system consisting of two ideal spheres or point sources can be directly analyzed within the framework of Newtonian $f(R)$ gravity, and they are generalizations of their counterparts (viz. conic sections) in Newtonian gravity. In addition, these results can also be effectively utilized in cases where the higher-order multipole moments of a body are negligible such as a planet orbiting the Sun. Under this circumstance, one can quantitatively analyze how the Sun's oblateness influences the planet's orbit. In light of these discussions, it can be concluded that these results constitute the foundation for studying numerous important phenomena  in the Solar system within $f(R)$ gravity. Besides, as discussed before, while the derivation assumes the bodies are composed of a perfect fluid, the obtained results in this paper also apply to solid bodies such as Earth-like planets. Thus, by applying these results, one can further explore the effects of the Solar multipole moments on Mercury's orbit and the precession of the planes of Earth-orbiting satellites in $f(R)$ gravity models.

This paper is organized as follows. In Sec.~\ref{Sec:second}, notation and related formulas in the STF formalism are described, and the metric $f(R)$ gravity is briefly reviewed. In Sec.~\ref{Sec:third}, the equations of motion for extended fluid bodies in an isolated self-gravitating system under the Newtonian limit of $f(R)$ gravity are derived. Furthermore, in Sec.~\ref{Sec:fourth}, the conserved quantities and spin dynamics in the system are discussed. In Sec.~\ref{Sec:fifth}, the orbital motion of a two-body system is explored in $f(R)$ gravity at the Newtonian order based on the results obtained in the previous sections. In Sec.~\ref{Sec:sixth}, the conclusions and the related discussions are presented. In the Appendix, the weak-field and slow-motion (WFSM) approximation of $f(R)$ gravity is revisited and the corresponding results are summarized. Throughout this paper, we use the international system of units, and the spacetime signature is $(-,+,+,+)$. The Greek letters $\alpha,\beta,\gamma,\cdots$ (running from 0 to 3) denote the spacetime indices, whereas the Latin letters $i,j,k,\cdots$ (running from 1 to 3) represent the spatial indices. The repeated Greek or Latin indices within a term mean that the sum should be taken over.
\section{Preliminary~\label{Sec:second}}
\subsection{Notation and related formulas in the STF formalism}
In the Newtonian limit of $f(R)$ gravity, the coordinates $(x^{\mu})=(ct,x_{i})$ behave like Minkowskian coordinates~\cite{Wu:2017huang}. With them, we further define $r=(x_{i}x_{i})^{1/2}$, $n_{i}=x_{i}/r$, $\partial_{\mu}=\partial/\partial x^{\mu}$, $\partial_{i}=\partial/\partial x_{i}$, $\boldsymbol{x}=x_{i}\partial_{i}$, $\Delta=\partial_{a}\partial_{a}$,
\begin{eqnarray}
\label{equ2.1}X_{I_{l}}&=&X_{i_{1}i_{2}\cdots i_{l}}:= x_{i_{1}}x_{i_{2}}\cdots x_{i_{l}},\\
\label{equ2.2}N_{I_{l}}&=&N_{i_{1}i_{2}\cdots i_{l}}:= n_{i_{1}}n_{i_{2}}\cdots n_{i_{l}},\\
\label{equ2.3}\partial_{I_{l}}&=&\partial_{i_{1}i_{2}\cdots i_{l}}:=\partial_{i_{1}}\partial_{i_{2}}\cdots\partial_{i_{l}},
\end{eqnarray}
where the spatial indices are raised or lowered using the Cartesian metric $\delta_{ij}=\delta^{ij}=\text{diag}(+1,+1,$ $+1)$.
The symbol $\epsilon_{ijk}$ with $\epsilon_{123}=1$ denotes the totally antisymmetric Levi-Civita tensor. For a Cartesian tensor $B_{I_{l}}=B_{i_{1}i_{2}\cdots i_{l}}$, its symmetric and STF parts~\cite{Thorne:1980ru} are, respectively,
\begin{eqnarray}
\label{equ2.4}&&S_{I_{l}}:=B_{(I_{l})}=B_{(i_{1}i_{2}\cdots i_{l})}:=\frac{1}{l!}\sum_{\sigma} B_{i_{\sigma(1)}i_{\sigma(2)}\cdots i_{\sigma(l)}},\\
\label{equ2.5}&&\hat{B}_{I_{l}}:=B_{\langle I_{l}\rangle}=B_{\langle i_{1}i_{2}\cdots i_{l}\rangle}
\displaystyle:=\sum_{k=0}^{\left[\frac{l}{2}\right]}b_{k}\delta_{(i_{1}i_{2}}\cdots\delta_{i_{2k-1}i_{2k}}S_{i_{2k+1}\cdots i_{l})a_{1}a_{1}\cdots a_{k}a_{k}},
\end{eqnarray}
where $\sigma$ runs over all permutations of $(12\cdots l)$, $\left[l/2\right]$ represents the integer part of $l/2$, and
\begin{equation}
\label{equ2.6}b_{k}:=(-1)^{k}\frac{(2l-2k-1)!!}{(2l-1)!!}\frac{l!}{(2k)!!(l-2k)!}.
\end{equation}
In the STF formalism, other related formulas of direct use in this paper are
\begin{eqnarray}
\label{equ2.7}&&\hat{N}_{I_{l}}=\sum_{k=0}^{[\frac{l}{2}]}b_{k}\delta_{(i_{1}i_{2}}\cdots\delta_{i_{2k-1}i_{2k}}
N_{i_{2k+1}\cdots i_{l})},\\
\label{equ2.8}&&\hat{\partial}_{I_{l}}=\sum_{k=0}^{\left[\frac{l}{2}\right]}b_{k}\delta_{(i_{1}i_{2}}\cdots\delta_{i_{2k-1}i_{2k}}
\partial_{i_{2k+1}\cdots i_{l})}\Delta^k,\\
\label{equ2.9}&&\hat{\partial}_{I_{l}}\left(\frac{F(r)}{r}\right)=\hat{N}_{I_{l}}\sum_{k=0}^{l}\frac{(l+k)!}{(-2)^{k}k!(l-k)!}
\frac{\partial_{r}^{l-k}F(r)}{r^{k+1}},\\
\label{equ2.9.5}&&\int\hat{N}_{I_{l}}\text{d}\varOmega=0\qquad \text{if}\qquad l\geqslant1,
\end{eqnarray}
where $\partial_r^{l-k}$ is the $(l-k)$-th derivative with respect to $r$, and $\text{d}\varOmega$ is the element of the solid angle about the radial vector.
\subsection{A brief review on $f(R)$ gravity}
Consider a spacetime with $g_{\mu\nu}$ as the metric, and let $g$ denote the metric determinant. In $f(R)$ gravity, the action is given by
\begin{equation}\label{equ2.10}
S=\frac{1}{2\kappa c}\int \text{d}^4x\sqrt{-g}f(R)+S_{M}(g^{\mu\nu},\psi),
\end{equation}
where $\kappa=8\pi G/c^{4}$ with $G$ as the gravitational constant, and $S_{M}(g^{\mu\nu},\psi)$ represents the matter action. Varying the above action with respect to the metric $g^{\mu\nu}$ can give rise to the gravitational field equations
\begin{equation}\label{equ2.11}
H_{\mu\nu}:=-\frac{1}{2}g_{\mu\nu}f+(R_{\mu\nu}+g_{\mu\nu}\square-\nabla_{\mu}\nabla_{\nu})\frac{\partial f}{\partial R}=\kappa T_{\mu\nu},
\end{equation}
where $T_{\mu\nu}$ is the energy-momentum tensor of matter fields. As in Ref.~\cite{Wu:2017huang}, $f(R)$ is assumed to have the polynomial form
\begin{equation}\label{equ2.12}
f(R)=R+a R^{2}+b R^{3}+\cdots.
\end{equation}
Here, a possible cosmological constant is ignored because we will adopt an expansion about a flat background spacetime under the WFSM approximation, and moreover, to link the above Lagrangian~(\ref{equ2.12}) to GR, the coefficient of $R$ is set to be $1$.
\section{Equations of motion for extended fluid bodies in an isolated self-gravitating system~\label{Sec:third}}
\subsection{Fundamental variables in the system }
We consider an isolated self-gravitating system consisting of $N$ extended fluid bodies, and the system satisfies the assumptions stated in Sec.~\ref{Sec:first}. Each body in the system is labeled by a unique index $A\in\{1, 2, \ldots, N\}$, and the volume of the body $A$ is represented by $V_{A}$. Let $\rho(t,\boldsymbol{x})$ be the mass-density field of the system, and because of $T^{00}(t,\boldsymbol{x})=\rho(t,\boldsymbol{x})c^2$ at the Newtonian order~\cite{Clifford2018,Eric2014}, according to Eqs.~(\ref{equA15}), (\ref{equA30}), (\ref{equA31}), and (\ref{equA36}),  the gravitational potential contributed by all the bodies in $f(R)$ gravity is
\begin{eqnarray}
\label{equ3.1}&&\varPhi(t,\boldsymbol{x})=\sum_{A=1}^{N}\varPhi^{A}(t,\boldsymbol{x}),
\end{eqnarray}
where
\begin{eqnarray}
\label{equ3.2}&&\varPhi^{A}(t,\boldsymbol{x})=U^{A}(t,\boldsymbol{x})+Y^{A}(t,\boldsymbol{x})
\end{eqnarray}
is the gravitational potential produced in part by body $A$ with
\begin{eqnarray}
\label{equ3.3}&&\displaystyle U^{A}(t,\boldsymbol{x})=\displaystyle G\int_{A}\frac{\rho(t,\boldsymbol{x}')}{|\boldsymbol{x}-\boldsymbol{x}'|}\text{d}^{3}x',\\
\label{equ3.4}&&\displaystyle Y^{A}(t,\boldsymbol{x})=\displaystyle \frac{G}{3}\int_{A}\frac{\rho(t,\boldsymbol{x}')\text{e}^{-m_{\text{s}}|\boldsymbol{x}-\boldsymbol{x}'|}}{|\boldsymbol{x}-\boldsymbol{x}'|}\text{d}^{3}x'
\end{eqnarray}
as the corresponding Coulomb-like and Yukawa-like potentials, respectively. Here, it should be noted that the parameter $m_{\text{s}}$ satisfying $m_{\text{s}}^{2}>0$ was introduced in Eq.~(\ref{equA8}), and the integrations in the above two equations need to be performed over the matter-filled volume $V_{A}$.
Within the framework of Newtonian $f(R)$ gravity (viz.~$f(R)$ gravity at the Newtonian order), inside the bodies, the Euler's equation
\begin{eqnarray}
\label{equ3.5}&&\rho\frac{\text{d}\boldsymbol{v}}{\text{d}t}=\rho\nabla\varPhi-\nabla p
\end{eqnarray}
and the continuity equation
\begin{eqnarray}
\label{equ3.6}&&\frac{\text{d}\rho}{\text{d}t}+\rho\nabla\cdot\boldsymbol{v}=0
\end{eqnarray}
govern the fluid dynamics, where $\nabla$ is the gradient operator, $\boldsymbol{v}(t,\boldsymbol{x})$ and $p(t,\boldsymbol{x})$ are the velocity field and the pressure field characterizing the fluid distribution of the system, and
\begin{eqnarray}
\label{equ3.7}&&\frac{\text{d}}{\text{d}t}:=\frac{\partial}{\partial t}+\boldsymbol{v}\cdot\nabla
\end{eqnarray}
is the convective time derivative that considers the intrinsic time variations brought about by the motion of the fluid elements. In metric $f(R)$ gravity, the covariant conservation of the matter energy-momentum tensor (which follows from diffeomorphism invariance) reduces at the Newtonian order to the continuity equation (\ref{equ3.6}). Consequently, the mass $m^{A}$ of each body $A$, defined by (\ref{equ3.8}), is constant in time.

For a body within the system, such as body $A$, its total mass is given by
\begin{eqnarray}
\label{equ3.8}&&m^{A}:=\int_{A}\rho(t,\boldsymbol{x})\,\text{d}^{3}x,
\end{eqnarray}
and its center-of-mass position is conventionally defined by
\begin{eqnarray}
\label{equ3.9}&&\boldsymbol{x}^{A}(t):=\frac{1}{m^{A}}\int_{A}\rho(t,\boldsymbol{x})\boldsymbol{x}\,\text{d}^{3}x.
\end{eqnarray}
With these definitions, the center-of-mass velocity and acceleration of body $A$ are expressed as
\begin{eqnarray}
\label{equ3.10}&&\boldsymbol{v}^{A}(t):=\frac{\text{d}\boldsymbol{x}^{A}}{\text{d}t}=\frac{1}{m^{A}}\int_{A}\rho(t,\boldsymbol{x})\boldsymbol{v}\,\text{d}^{3}x
\end{eqnarray}
and
\begin{eqnarray}
\label{equ3.11}&&\boldsymbol{a}^{A}(t):=\frac{\text{d}\boldsymbol{v}^{A}}{\text{d}t}=\frac{1}{m^{A}}\int_{A}\rho(t,\boldsymbol{x})\frac{\text{d}\boldsymbol{v}}{\text{d}t}\,\text{d}^{3}x,
\end{eqnarray}
respectively, where the derivations of these expressions rely on the following useful formula relevant to the convective time derivative~\cite{Eric2014},
\begin{eqnarray}
\label{equ3.12}&&\frac{\text{d}}{\text{d}t}\int_{A}\rho(t,\boldsymbol{x})F(t,\boldsymbol{x})\text{d}^{3}x=\int_{A}\rho(t,\boldsymbol{x})\frac{\text{d}F}{\text{d}t}\text{d}^{3}x.
\end{eqnarray}
The core goal of the coarse-grained description of the inter-body dynamics for the system in the Newtonian limit of $f(R)$ gravity is to obtain the equations of motion of the bodies and recast these equations in terms of their multipole moments.
Therefore, how to perform a multipole analysis on the equations of motion of the bodies by applying the STF formalism  in terms of irreducible Cartesian tensors is the central task of this paper.
\subsection{The internal and external potentials}
By just plugging the gravitational potential (\ref{equ3.1}) contributed by all the bodies in the system and the Euler's equation (\ref{equ3.5}) in the gravitational field into the expression for the center-of-mass acceleration~(\ref{equ3.11}) of body $A$, its equation of motion is derived,
\begin{eqnarray}
\label{equ3.13}&&m^{A}\boldsymbol{a}^{A}=\int_{A}\rho\nabla\varPhi\,\text{d}^{3}x,
\end{eqnarray}
where the identity
\begin{eqnarray}
\label{equ3.14}&&\int_{A}\nabla p\,\text{d}^{3}x=\oint_{\partial A}p\text{d}\boldsymbol{S}=0
\end{eqnarray}
is employed because of Gauss's theorem and the fact that the pressure vanishes on the boundary $\partial A$ of the integration region. If the internal potential $\varPhi^{A}(t,\boldsymbol{x})$ is defined as the potential generated by body $A$ and the external potential $\varPhi^{-A}(t,\boldsymbol{x})$
is defined as the potential contributed by all other bodies, the gravitational potential $\varPhi(t,\boldsymbol{x})$ contributed by all the bodies should have the following decomposition, namely
\begin{eqnarray}
\label{equ3.15}&&\varPhi=\varPhi^{A}+\varPhi^{-A}.
\end{eqnarray}
With this decomposition, the equation of motion of body $A$ can be reexpressed as
\begin{eqnarray}
\label{equ3.16}&&m^{A}\boldsymbol{a}^{A}=\int_{A}\rho\nabla\varPhi^{A}\,\text{d}^{3}x+\int_{A}\rho\nabla\varPhi^{-A}\,\text{d}^{3}x.
\end{eqnarray}
Similar to the corresponding result in GR, we can also prove that the internal potential $\varPhi^{A}$ makes no contribution to the equation of motion of body $A$. Obviously, from Eq.~(\ref{equ3.2}), we only need to show
\begin{eqnarray}
\label{equ3.17}&&\int_{A}\rho\nabla\varPhi^{A}\,\text{d}^{3}x=\int_{A}\rho\nabla U^{A}\,\text{d}^{3}x+\int_{A}\rho\nabla Y^{A}\,\text{d}^{3}x=0.\qquad
\end{eqnarray}
Considering the $x_{i}$ components of the two terms on the right side of above equation and substituting the expressions of the Coulomb-like potential~(\ref{equ3.3}) and Yukawa-like potential~(\ref{equ3.4}), we have that
\begin{eqnarray}
\label{equ3.18}\int_{A}\rho\partial_{i}U^{A}\,\text{d}^{3}x&=&G\int_{A}\int_{A}\rho\rho'\frac{\partial}{\partial x_{i}}\left(\frac{1}{|\boldsymbol{x}-\boldsymbol{x}'|}\right)\,\text{d}^{3}x\text{d}^{3}x'\notag\\
&=&G\int_{A}\int_{A}\rho'\rho\frac{\partial}{\partial x'_{i}}\left(\frac{1}{|\boldsymbol{x}'-\boldsymbol{x}|}\right)\,\text{d}^{3}x'\text{d}^{3}x\notag\\
&=&-G\int_{A}\int_{A}\rho\rho'\frac{\partial}{\partial x_{i}}\left(\frac{1}{|\boldsymbol{x}-\boldsymbol{x}'|}\right)\,\text{d}^{3}x\text{d}^{3}x'
\end{eqnarray}
and
\begin{eqnarray}
\label{equ3.19}\int_{A}\rho\partial_{i}Y^{A}\,\text{d}^{3}x&=&\frac{G}{3}\int_{A}\int_{A}\rho\rho'\frac{\partial}{\partial x_{i}}\left(\frac{\text{e}^{-m_{\text{s}}|\boldsymbol{x}-\boldsymbol{x}'|}}{|\boldsymbol{x}-\boldsymbol{x}'|}\right)\,\text{d}^{3}x\text{d}^{3}x'\notag\\
&=&\frac{G}{3}\int_{A}\int_{A}\rho'\rho\frac{\partial}{\partial x'_{i}}\left(\frac{\text{e}^{-m_{\text{s}}|\boldsymbol{x}'-\boldsymbol{x}|}}{|\boldsymbol{x}'-\boldsymbol{x}|}\right)\,\text{d}^{3}x'\text{d}^{3}x\notag\\
&=&-\frac{G}{3}\int_{A}\int_{A}\rho\rho'\frac{\partial}{\partial x_{i}}\left(\frac{\text{e}^{-m_{\text{s}}|\boldsymbol{x}-\boldsymbol{x}'|}}{|\boldsymbol{x}-\boldsymbol{x}'|}\right)\,\text{d}^{3}x\text{d}^{3}x',\qquad
\end{eqnarray}
where $\rho'$ is the mass density expressed in terms of $t$ and $\boldsymbol{x}'$. The derivations of the two identities are similar. In each derivation, the variables of integration in the second step are swapped (which acts as a mere relabeling of dummy variables and introduces no sign change), and the minus in the third step originates from the fact that both $1/|\boldsymbol{x}-\boldsymbol{x}'|$ and $\text{e}^{-m_{\text{s}}|\boldsymbol{x}-\boldsymbol{x}'|}/|\boldsymbol{x}-\boldsymbol{x}'|$ depend on the difference between $\boldsymbol{x}$ and $\boldsymbol{x}'$. With these two identities, one could find that Eq.~(\ref{equ3.17}) holds, and as a result,
the equation of motion of body $A$ becomes
\begin{eqnarray}
\label{equ3.20}&&m^{A}\boldsymbol{a}^{A}=\int_{A}\rho\nabla\varPhi^{-A}\,\text{d}^{3}x,
\end{eqnarray}
which confirms that the motion of body $A$ is determined solely by the external potential generated by other bodies.
\subsection{Multipole analysis on $1/|\boldsymbol{x}-\boldsymbol{x}'|$ and $\text{e}^{-m_{\text{s}}|\boldsymbol{x}-\boldsymbol{x}'|}/|\boldsymbol{x}-\boldsymbol{x}'|$}
Equation~(\ref{equ3.2}) indicates that the gravitational potential generated by each body within the system comprises the Coulomb-like and Yukawa-like potentials. Thus, by definition,
for body $A$, the external potential $\varPhi^{-A}(t,\boldsymbol{x})$ could be expressed as the sum of the external Coulomb-like potential $U^{-A}(t,\boldsymbol{x})$ and the external Yukawa-like potential $Y^{-A}(t,\boldsymbol{x})$,
\begin{eqnarray}
\label{equ3.21}&&\varPhi^{-A}=U^{-A}+Y^{-A},
\end{eqnarray}
where from Eqs.~(\ref{equ3.3}) and~(\ref{equ3.4}),
\begin{eqnarray}
\label{equ3.22}&&U^{-A}(t,\boldsymbol{x}):=\sum_{B\neq A}U^{B}(t,\boldsymbol{x})=\sum_{B\neq A}G\int_{B}\frac{\rho(t,\boldsymbol{x}')}{|\boldsymbol{x}-\boldsymbol{x}'|}\text{d}^{3}x',\\
\label{equ3.23}&&Y^{-A}(t,\boldsymbol{x}):=\sum_{B\neq A}Y^{B}(t,\boldsymbol{x})=\sum_{B\neq A}\frac{G}{3}\int_{B}\frac{\rho(t,\boldsymbol{x}')\text{e}^{-m_{\text{s}}|\boldsymbol{x}-\boldsymbol{x}'|}}{|\boldsymbol{x}-\boldsymbol{x}'|}\text{d}^{3}x'
\end{eqnarray}
are, respectively, the Coulomb-like and Yukawa-like potentials produced by all other bodies. Since the derivatives of these two external potentials with respect to the spatial coordinates play crucial roles in the equation of motion of body $A$, we also present their expressions here,
\begin{eqnarray}
\label{equ3.24}&&\partial_{i}U^{-A}=\sum_{B\neq A}G\int_{B}\rho'\frac{\partial}{\partial x_{i}}\left(\frac{1}{|\boldsymbol{x}-\boldsymbol{x}'|}\right)\text{d}^{3}x',\\
\label{equ3.25}&&\partial_{i}Y^{-A}=\sum_{B\neq A}\frac{G}{3}\int_{B}\rho'\frac{\partial}{\partial x_{i}}\left(\frac{\text{e}^{-m_{\text{s}}|\boldsymbol{x}-\boldsymbol{x}'|}}{|\boldsymbol{x}-\boldsymbol{x}'|}\right)\text{d}^{3}x'.
\end{eqnarray}
To analyze the external potentials and their derivatives by applying the STF formalism and investigate the system's dynamical equations in the following subsections, we will now perform multipole expansions for $1/|\boldsymbol{x}-\boldsymbol{x}'|$, $\text{e}^{-m_{\text{s}}|\boldsymbol{x}-\boldsymbol{x}'|}/|\boldsymbol{x}-\boldsymbol{x}'|$, and their derivatives by invoking the assumption that the typical size of each body is far less than the typical separation between bodies.

It is shown from Eqs.~(\ref{equ3.20}), (\ref{equ3.22}), and (\ref{equ3.23})  that the vectors $\boldsymbol{x}$ and $\boldsymbol{x}'$ vary within the characteristic sizes of body $A$ and body $B$, respectively. Since the bodies in the system are well separated, and their characteristic sizes are far less than the separations between them, there is $|\boldsymbol{x}-\boldsymbol{x}^{A}|\ll|\boldsymbol{x}'-\boldsymbol{x}^{A}|$, where $\boldsymbol{x}^{A}$ is the center-of-mass position vector of body $A$. In order to separate the vector $\boldsymbol{x}-\boldsymbol{x}^{A}$ from the vector $\boldsymbol{x}'-\boldsymbol{x}^{A}$, the multipole expansions for $1/|\boldsymbol{x}-\boldsymbol{x}'|$ and $\text{e}^{-m_{\text{s}}|\boldsymbol{x}-\boldsymbol{x}'|}/|\boldsymbol{x}-\boldsymbol{x}'|$ in the center-of-mass coordinate system of body $A$ need to be performed first. Taking this into account, $|\boldsymbol{x}-\boldsymbol{x}'|$ should be transformed into $|(\boldsymbol{x}-\boldsymbol{x}^{A})-(\boldsymbol{x}'-\boldsymbol{x}^{A})|$, and then,  $1/|\boldsymbol{x}-\boldsymbol{x}'|$ and $\text{e}^{-m_{\text{s}}|\boldsymbol{x}-\boldsymbol{x}'|}/|\boldsymbol{x}-\boldsymbol{x}'|$ can be rewritten as
\begin{eqnarray}
\label{equ3.26}&&\frac{1}{|\boldsymbol{x}-\boldsymbol{x}'|}=\frac{1}{|\boldsymbol{y}-\boldsymbol{y}'|},\\
\label{equ3.27}&&\frac{\text{e}^{-m_{\text{s}}|\boldsymbol{x}-\boldsymbol{x}'|}}{|\boldsymbol{x}-\boldsymbol{x}'|}=\frac{\text{e}^{-m_{\text{s}}|\boldsymbol{y}-\boldsymbol{y}'|}}{|\boldsymbol{y}-\boldsymbol{y}'|}
\end{eqnarray}
with $\boldsymbol{y}:=\boldsymbol{x}-\boldsymbol{x}^{A}$ and $\boldsymbol{y}':=\boldsymbol{x}'-\boldsymbol{x}^{A}$. One is able to perform the multipole expansions for $1/|\boldsymbol{y}-\boldsymbol{y}'|$ and $\text{e}^{-m_{\text{s}}|\boldsymbol{y}-\boldsymbol{y}'|}/|\boldsymbol{y}-\boldsymbol{y}'|$ with the help of the following two formulas~\cite{Wu:2017huang,Wu:2022akq}
\begin{eqnarray}
\label{equ3.28}&&\frac{1}{|\boldsymbol{y}-\boldsymbol{y}'|}=\sum_{l=0}^{\infty}\frac{(2l-1)!!}{l!}\frac{(r_< )^l}{(r_> )^{l+1}}\hat{N}_{I_{l}}(\theta,\varphi)\hat{N}_{I_{l}}(\theta',\varphi'),\\
\label{equ3.29}&&\frac{\text{e}^{-m_{\text{s}}|\boldsymbol{y}-\boldsymbol{y}'|}}{|\boldsymbol{y}-\boldsymbol{y}'|}=\sum_{l=0}^{\infty}\frac{(2l+1)!!}{l!}m_{\text{s}}\text{i}_{l}(m_{\text{s}}r_<)\text{k}_{l}(m_{\text{s}}r_> )\hat{N}_{I_{l}}(\theta,\varphi)\hat{N}_{I_{l}}(\theta',\varphi').
\end{eqnarray}
In the two formulas, $r_{<}$ represents the lesser of $r=|\boldsymbol{y}|$ and $r'=|\boldsymbol{y}'|$, and $r_{>}$ the greater. The components of the vectors $\boldsymbol{y}$ and $\boldsymbol{y}'$ are denoted by
\begin{eqnarray}
\label{equ3.30}&&y_{i}=x_{i}-x^{A}_{i}=:\left(x-x^{A}\right)_{i},\\
\label{equ3.31}&&y'_{i}=x'_{i}-x^{A}_{i}=:\left(x'-x^{A}\right)_{i},
\end{eqnarray}
their angle coordinates are given by $(\theta,\varphi)$ and $(\theta',\varphi')$, $N_{I_{l}}(\theta,\varphi)=n_{i_{1}}n_{i_{2}}\cdots n_{i_{l}}$ with $n_{i}:=y_{i}/r=y_{i}/(y_{j}y_{j})^{1/2}$, and $N'_{I_{l}}(\theta',\varphi')=n'_{i_{1}}n'_{i_{2}}\cdots n'_{i_{l}}$ with $n'_{i}:=y'_{i}/r'=y'_{i}/(y'_{j}y'_{j})^{1/2}$. In addition, in Eq.~(\ref{equ3.29}),
 \begin{eqnarray}
\label{equ3.32}\text{i}_{l}(z):=\sqrt{\frac{\pi}{2z}}\text{I}_{l+\frac{1}{2}}(z),\qquad \text{k}_{l}(z):=\sqrt{\frac{2}{\pi z}}\text{K}_{l+\frac{1}{2}}(z)
\end{eqnarray}
are the spherical modified Bessel functions of $l$-order~\cite{Arfken1985} with
$\text{I}_{l+1/2}(z)$, $\text{K}_{l+1/2}(z)$ as the modified Bessel functions of $(l+1/2)$-order. Continuing along previous lines and directly inserting Eqs.~(\ref{equ3.28}) and (\ref{equ3.29}) into Eqs.~(\ref{equ3.26}) and (\ref{equ3.27}), the multipole expansions of $1/|\boldsymbol{x}-\boldsymbol{x}'|$ and $\text{e}^{-m_{\text{s}}|\boldsymbol{x}-\boldsymbol{x}'|}/|\boldsymbol{x}-\boldsymbol{x}'|$ in the center-of-mass coordinate system of body $A$ are obtained,
\begin{eqnarray}
\label{equ3.33}&&\frac{1}{|\boldsymbol{x}-\boldsymbol{x}'|}=\sum_{l=0}^{\infty}\frac{(2l-1)!!}{l!}\frac{r^l}{r'^{l+1}}\hat{N}_{I_{l}}(\theta,\varphi)\hat{N}_{I_{l}}(\theta',\varphi'),\\
\label{equ3.34}&&\frac{\text{e}^{-m_{\text{s}}|\boldsymbol{x}-\boldsymbol{x}'|}}{|\boldsymbol{x}-\boldsymbol{x}'|}=\sum_{l=0}^{\infty}\frac{(2l+1)!!}{l!}m_{\text{s}}\text{i}_{l}(m_{\text{s}}r)\text{k}_{l}(m_{\text{s}}r')\hat{N}_{I_{l}}(\theta,\varphi)\hat{N}_{I_{l}}(\theta',\varphi').
\end{eqnarray}
Both expansions need to be further transformed into a more convenient form for use. On the basis of definitions and Eqs.~(\ref{equ2.7})--(\ref{equ2.9}), the two identities
\begin{eqnarray}
\label{equ3.35}&&r^l\hat{N}_{I_{l}}(\theta,\varphi)=\left(x-x^{A}\right)_{\langle I_{l}\rangle},\\
\label{equ3.36}&&\hat{\partial'}_{I_{l}}\left(\frac{1}{r'}\right)=\frac{(-1)^{l}(2l-1)!!}{r'^{l+1}}\hat{N}_{I_{l}}(\theta',\varphi')
\end{eqnarray}
are first established, where $\left(x-x^{A}\right)_{I_{l}}:=\left(x-x^{A}\right)_{i_{1}}\cdots\left(x-x^{A}\right)_{i_{l}}$ and $\partial'_{I_{l}}=\partial'_{i_{1}}\cdots\partial'_{i_{l}}$ with $\partial'_{i}=\partial/\partial x'_{i}$. When introducing $\partial^{A}_{i}:=\partial/\partial x_{i}^{A}$, the identity
\begin{eqnarray}
\label{equ3.37}\partial^{A}_{i}\left(\frac{1}{|\boldsymbol{x}^{A}-\boldsymbol{x}'|}\right)=-\partial'_{i}\left(\frac{1}{r'}\right)
\end{eqnarray}
also holds. By means of these three identities, the expansion~(\ref{equ3.33}) can immediately be recast into
\begin{eqnarray}
\label{equ3.38}\frac{1}{|\boldsymbol{x}-\boldsymbol{x}'|}&=&\sum_{l=0}^{\infty}\frac{1}{l!}\left(x-x^{A}\right)_{\langle I_{l}\rangle}\,\partial^{A}_{I_{l}}\left(\frac{1}{|\boldsymbol{x}^{A}-\boldsymbol{x}'|}\right),
\end{eqnarray}
and here, $\partial^{A}_{I_{l}}=\partial^{A}_{i_{1}}\partial^{A}_{i_{2}}\cdots\partial^{A}_{i_{l}}$. As to the handling of the expansion~(\ref{equ3.34}), the situation becomes complicated. As shown in Refs.~\cite{Arfken1985,Abramowitz1965}, the spherical modified Bessel functions of $l$-order have the following expressions,
\begin{equation}\label{equ3.39}
\left\{\begin{array}{lll}
\displaystyle \text{i}_{l}(z)&\displaystyle=z^l\left(\frac{\text{d}}{z\text{d}z}\right)^{l}\left(\frac{\sinh{z}}{z}\right)\smallskip\\
\displaystyle \phantom{\text{i}_{l}(z)}&\displaystyle=\frac{\text{e}^{z}}{2z}\left(\sum_{k=0}^{l}\frac{(-1)^{k}(l+k)!}{k!(l-k)!}\frac{1}{(2z)^{k}}\right)-(-1)^{l}\frac{\text{e}^{-z}}{2z}\left(\sum_{k=0}^{l}\frac{(l+k)!}{k!(l-k)!}\frac{1}{(2z)^{k}}\right),\\
\displaystyle \text{k}_{l}(z)&=\displaystyle (-1)^{l}z^{l}\left(\frac{\text{d}}{z\text{d}z}\right)^{l}\frac{\text{e}^{-z}}{z}=\frac{\text{e}^{-z}}{z}\sum_{k=0}^{l}\frac{(l+k)!}{k!(l-k)!}\frac{1}{(2z)^{k}}.
\end{array}\right.
\end{equation}
If we define the function
\begin{eqnarray}
\label{equ3.40}&&\updelta_{l}(z):=(2l+1)!!\bigg(\frac{\text{d}}{z\text{d}z}\bigg)^{l}\bigg(\frac{\sinh{z}}{z}\bigg),
\end{eqnarray}
there is
\begin{eqnarray}
\label{equ3.41}\displaystyle &&\text{i}_{l}(m_{\text{s}}r)=\frac{(m_{\text{s}}r)^{l}}{(2l+1)!!}\updelta_{l}(m_{\text{s}}r).
\end{eqnarray}
Moreover, by use of Eqs.~(\ref{equ2.9}), (\ref{equ3.39}), and
\begin{eqnarray}
\label{equ3.42}\text{e}^{-z}=(-1)^{l-k}\frac{\text{d}^{l-k}}{\text{d}z^{l-k}}\text{e}^{-z},
\end{eqnarray}
we can infer that
\begin{eqnarray}
\label{equ3.43}&&\text{k}_{l}(m_{\text{s}}r')\hat{N}_{I_{l}}(\theta',\varphi')=\frac{(-1)^{l}}{m_{\text{s}}^{l+1}}\hat\partial'_{I_{l}}\left(\frac{\text{e}^{-m_{\text{s}}r'}}{r'}\right).
\end{eqnarray}
Equations~(\ref{equ3.35}),  (\ref{equ3.41}), and (\ref{equ3.43}) collectively suggest that the expansion~(\ref{equ3.34}) should be recast into
\begin{eqnarray}
\label{equ3.44}\frac{\text{e}^{-m_{\text{s}}|\boldsymbol{x}-\boldsymbol{x}'|}}{|\boldsymbol{x}-\boldsymbol{x}'|}&=&\sum_{l=0}^{\infty}\frac{1}{l!}\updelta_{l}\left(m_{\text{s}}|\boldsymbol{x}-\boldsymbol{x}^{A}|\right)\left(x-x^{A}\right)_{\langle I_{l}\rangle}\,\partial^{A}_{I_{l}}\left(\frac{\text{e}^{-m_{\text{s}}|\boldsymbol{x}^{A}-\boldsymbol{x}'|}}{|\boldsymbol{x}^{A}-\boldsymbol{x}'|}\right).
\end{eqnarray}
Clearly, after performing the multipole expansions for $1/|\boldsymbol{x}-\boldsymbol{x}'|$ and $\text{e}^{-m_{\text{s}}|\boldsymbol{x}-\boldsymbol{x}'|}/|\boldsymbol{x}-\boldsymbol{x}'|$ in the center-of-mass coordinate system of body $A$,  the vector $\boldsymbol{x}-\boldsymbol{x}^{A}$ is completely separated from the vector $\boldsymbol{x}'-\boldsymbol{x}^{A}$. Given that $\boldsymbol{x}-\boldsymbol{x}^{A}$ represents the position vector of a point within the body $A$ relative to its center-of-mass, the parts associated with $\boldsymbol{x}-\boldsymbol{x}^{A}$
in expansions~(\ref{equ3.38}) and (\ref{equ3.44}) could be employed to define the multipole moments of body $A$. It can be anticipated that the multipole moments of body $B$ should also be solely defined, and to achieve this purpose, the vector $\boldsymbol{x}'-\boldsymbol{x}^{B}$ must also be separated by performing the further multipole expansions for $1/|\boldsymbol{x}^{A}-\boldsymbol{x}'|$ and $\text{e}^{-m_{\text{s}}|\boldsymbol{x}^{A}-\boldsymbol{x}'|}/|\boldsymbol{x}^{A}-\boldsymbol{x}'|$ in the center-of-mass coordinate system of body $B$.

Remember that the vector $\boldsymbol{x}'$ varies within the characteristic size of body $B$, and by again invoking the assumption that the characteristic size of body $B$ is far less than the separation between bodies $A$ and $B$,
there ought to be $|\boldsymbol{x}'-\boldsymbol{x}^{B}|\ll|\boldsymbol{x}^{A}-\boldsymbol{x}^{B}|$. Based on this conclusion, $|\boldsymbol{x}^{A}-\boldsymbol{x}'|$ can be transformed into $|\boldsymbol{x}^{AB}-\boldsymbol{\bar{x}}'|$ with $\boldsymbol{x}^{AB}:=\boldsymbol{x}^{A}-\boldsymbol{x}^{B}$ and $\boldsymbol{\bar{x}}':=\boldsymbol{x}'-\boldsymbol{x}^{B}$. Then, starting from formulas~(\ref{equ3.28}) and (\ref{equ3.29}) again, one only needs to repeat the previous derivation processes so that the multipole expansions for $1/|\boldsymbol{x}^{A}-\boldsymbol{x}'|$ and $\text{e}^{-m_{\text{s}}|\boldsymbol{x}^{A}-\boldsymbol{x}'|}/|\boldsymbol{x}^{A}-\boldsymbol{x}'|$ in the center-of-mass coordinate system of body $B$ can be obtained,
\begin{eqnarray}
\label{equ3.45}&&\frac{1}{|\boldsymbol{x}^{A}-\boldsymbol{x}'|}\hspace{-0.05cm}=\hspace{-0.05cm}\frac{1}{|\boldsymbol{x}^{AB}-\boldsymbol{\bar{x}}'|}\hspace{-0.05cm}=\hspace{-0.1cm}\sum_{l'=0}^{\infty}\frac{(-1)^{l'}}{l'!}\left(x'-x^{B}\right)_{\langle J_{l'}\rangle}\hspace{-0.05cm}\partial^{A}_{J_{l'}}\left(\frac{1}{|\boldsymbol{x}^{A}-\boldsymbol{x}^{B}|}\right),\\
\label{equ3.46}&&\frac{\text{e}^{-m_{\text{s}}|\boldsymbol{x}^{A}-\boldsymbol{x}'|}}{|\boldsymbol{x}^{A}-\boldsymbol{x}'|}\hspace{-0.05cm}=\hspace{-0.05cm}\frac{\text{e}^{-m_{\text{s}}|\boldsymbol{x}^{AB}-\boldsymbol{\bar{x}}'|}}{|\boldsymbol{x}^{AB}-\boldsymbol{\bar{x}}'|}=\sum_{l'=0}^{\infty}\frac{(-1)^{l'}}{l'!}\updelta_{l'}\left(m_{\text{s}}|\boldsymbol{x}'-\boldsymbol{x}^{B}|\right)\left(x'-x^{B}\right)_{\langle J_{l'}\rangle}\hspace{-0.05cm}\partial^{A}_{J_{l'}}\left(\frac{\text{e}^{-m_{\text{s}}|\boldsymbol{x}^{A}-\boldsymbol{x}^{B}|}}{|\boldsymbol{x}^{A}-\boldsymbol{x}^{B}|}\right),\qquad
\end{eqnarray}
where $(x'-x^{B})_{J_{l'}}:=(x'-x^{B})_{j_{1}}\cdots(x'-x^{B})_{j_{l'}}$ with $(x'-x^{B})_{j}$ as the components of the vector $\boldsymbol{x}'-\boldsymbol{x}^{B}$ and $\partial^{A}_{J_{l'}}=\partial^{A}_{j_{1}}\partial^{A}_{j_{2}}\cdots\partial^{A}_{j_{l'}}$. After deriving these two results, by replacing them into Eqs.~(\ref{equ3.38}) and (\ref{equ3.44}) , the final multipole expansions for $1/|\boldsymbol{x}-\boldsymbol{x}'|$ and $\text{e}^{-m_{\text{s}}|\boldsymbol{x}-\boldsymbol{x}'|}/|\boldsymbol{x}-\boldsymbol{x}'|$ are achieved,
\begin{eqnarray}
\label{equ3.47}&&\frac{1}{|\boldsymbol{x}-\boldsymbol{x}'|}=\sum_{l=0}^{\infty}\sum_{l'=0}^{\infty}\frac{(-1)^{l'}}{l!\,l'!}\left(x-x^{A}\right)_{\langle I_{l}\rangle}\left(x'-x^{B}\right)_{\langle J_{l'}\rangle}\,\partial^{A}_{I_{l}J_{l'}}\left(\frac{1}{|\boldsymbol{x}^{A}-\boldsymbol{x}^{B}|}\right),\\
\label{equ3.48}&&\frac{\text{e}^{-m_{\text{s}}|\boldsymbol{x}-\boldsymbol{x}'|}}{|\boldsymbol{x}-\boldsymbol{x}'|}=\sum_{l=0}^{\infty}\sum_{l'=0}^{\infty}\frac{(-1)^{l'}}{l!\,l'!}\updelta_{l}\left(m_{\text{s}}|\boldsymbol{x}-\boldsymbol{x}^{A}|\right)\left(x-x^{A}\right)_{\langle I_{l}\rangle}\notag\\
&&\phantom{\frac{\text{e}^{-m_{\text{s}}|\boldsymbol{x}-\boldsymbol{x}'|}}{|\boldsymbol{x}-\boldsymbol{x}'|}=}\times\updelta_{l'}\left(m_{\text{s}}|\boldsymbol{x}'-\boldsymbol{x}^{B}|\right)\left(x'-x^{B}\right)_{\langle J_{l'}\rangle}\partial^{A}_{I_{l}J_{l'}}\left(\frac{\text{e}^{-m_{\text{s}}|\boldsymbol{x}^{A}-\boldsymbol{x}^{B}|}}{|\boldsymbol{x}^{A}-\boldsymbol{x}^{B}|}\right).
\end{eqnarray}
The above two equations present the appropriate forms of the multipole expansions for $1/|\boldsymbol{x}-\boldsymbol{x}'|$ and $\text{e}^{-m_{\text{s}}|\boldsymbol{x}-\boldsymbol{x}'|}/|\boldsymbol{x}-\boldsymbol{x}'|$ in applications.
The same methodology can also be employed to determine the corresponding forms of their derivatives with respect to coordinates. From Eqs.~(\ref{equ3.38}) and (\ref{equ3.44}), we directly deduce that
\begin{eqnarray}
\label{equ3.49}&&\frac{\partial}{\partial x_{i}}\left(\frac{1}{|\boldsymbol{x}-\boldsymbol{x}'|}\right)=\sum_{l=0}^{\infty}\frac{1}{l!}\left(x-x^{A}\right)_{\langle I_{l}\rangle}\,\partial^{A}_{iI_{l}}\left(\frac{1}{|\boldsymbol{x}^{A}-\boldsymbol{x}'|}\right),\\
\label{equ3.50}&&\frac{\partial}{\partial x_{i}}\left(\frac{\text{e}^{-m_{\text{s}}|\boldsymbol{x}-\boldsymbol{x}'|}}{|\boldsymbol{x}-\boldsymbol{x}'|}\right)=\sum_{l=0}^{\infty}\frac{1}{l!}\updelta_{l}\left(m_{\text{s}}|\boldsymbol{x}-\boldsymbol{x}^{A}|\right)\left(x-x^{A}\right)_{\langle I_{l}\rangle}\,\partial^{A}_{iI_{l}}\left(\frac{\text{e}^{-m_{\text{s}}|\boldsymbol{x}^{A}-\boldsymbol{x}'|}}{|\boldsymbol{x}^{A}-\boldsymbol{x}'|}\right),\qquad
\end{eqnarray}
and then, inserting the expansions~(\ref{equ3.45}) and (\ref{equ3.46}) into the above two equations gives rise to the corresponding forms of the multipole expansions for the derivatives of $1/|\boldsymbol{x}-\boldsymbol{x}'|$ and $\text{e}^{-m_{\text{s}}|\boldsymbol{x}-\boldsymbol{x}'|}/|\boldsymbol{x}$ $-\boldsymbol{x}'|$,
\begin{eqnarray}
\label{equ3.51}&&\frac{\partial}{\partial x_{i}}\left(\frac{1}{|\boldsymbol{x}-\boldsymbol{x}'|}\right)=\sum_{l=0}^{\infty}\sum_{l'=0}^{\infty}\frac{(-1)^{l'}}{l!\,l'!}\left(x-x^{A}\right)_{\langle I_{l}\rangle}\left(x'-x^{B}\right)_{\langle J_{l'}\rangle}\partial^{A}_{iI_{l}J_{l'}}\left(\frac{1}{|\boldsymbol{x}^{A}-\boldsymbol{x}^{B}|}\right),\quad\qquad\\
\label{equ3.52}&&\frac{\partial}{\partial x_{i}}\left(\frac{\text{e}^{-m_{\text{s}}|\boldsymbol{x}-\boldsymbol{x}'|}}{|\boldsymbol{x}-\boldsymbol{x}'|}\right)=\sum_{l=0}^{\infty}\sum_{l'=0}^{\infty}\frac{(-1)^{l'}}{l!\,l'!}\updelta_{l}\left(m_{\text{s}}|\boldsymbol{x}-\boldsymbol{x}^{A}|\right)\left(x-x^{A}\right)_{\langle I_{l}\rangle}\notag\\
&&\phantom{\frac{\partial}{\partial x_{i}}\left(\frac{\text{e}^{-m_{\text{s}}|\boldsymbol{x}-\boldsymbol{x}'|}}{|\boldsymbol{x}-\boldsymbol{x}'|}\right)=}\times\updelta_{l'}\left(m_{\text{s}}|\boldsymbol{x}'-\boldsymbol{x}^{B}|\right)\left(x'-x^{B}\right)_{\langle J_{l'}\rangle}\partial^{A}_{iI_{l}J_{l'}}\left(\frac{\text{e}^{-m_{\text{s}}|\boldsymbol{x}^{A}-\boldsymbol{x}^{B}|}}{|\boldsymbol{x}^{A}-\boldsymbol{x}^{B}|}\right).
\end{eqnarray}
\subsection{Equations of motion for bodies in the system}
After the hard work in the previous subsection, we are in a position to discuss the equations of motion for bodies in the system. For body $A$, Eq.~(\ref{equ3.20}) indicates that its motion is only affected by the gradient of the external potential, namely $\nabla\varPhi^{-A}$, and thus, by plugging Eqs.~(\ref{equ3.21}),  (\ref{equ3.24}), and (\ref{equ3.25}),  the equation of motion of body $A$ can be explicitly expressed as
\begin{eqnarray}
\label{equ3.53}m^{A}a_{i}^{A}&=&G\sum_{B\neq A}\int_{A}\int_{B}\rho\rho'\frac{\partial}{\partial x_{i}}\left(\frac{1}{|\boldsymbol{x}-\boldsymbol{x}'|}\right)\,\text{d}^{3}x\text{d}^{3}x'+\frac{G}{3}\sum_{B\neq A}\int_{A}\int_{B}\rho\rho'\frac{\partial}{\partial x_{i}}\left(\frac{\text{e}^{-m_{\text{s}}|\boldsymbol{x}-\boldsymbol{x}'|}}{|\boldsymbol{x}-\boldsymbol{x}'|}\right)\,\text{d}^{3}x\text{d}^{3}x'.
\end{eqnarray}
In view that the multipole expansions for the spatial derivatives of $1/|\boldsymbol{x}-\boldsymbol{x}'|$ and $\text{e}^{-m_{\text{s}}|\boldsymbol{x}-\boldsymbol{x}'|}/|\boldsymbol{x}$ $-\boldsymbol{x}'|$ have been provided in Eqs.~(\ref{equ3.51}) and (\ref{equ3.52}), we are able to rewrite the equation of motion of body $A$ as
\begin{eqnarray}
\label{equ3.54}m^{A}a_{i}^{A}&=&G\sum_{B\neq A}\sum_{l=0}^{\infty}\sum_{l'=0}^{\infty}\frac{(-1)^{l'}}{l!\,l'!}\hat{M}^{A}_{I_{l}}(t)\hat{M}^{B}_{J_{l'}}(t)\partial^{A}_{iI_{l}J_{l'}}\left(\frac{1}{|\boldsymbol{x}^{A}-\boldsymbol{x}^{B}|}\right)\notag\\
&&+\frac{G}{3}\sum_{B\neq A}\sum_{l=0}^{\infty}\sum_{l'=0}^{\infty}\frac{(-1)^{l'}}{l!\,l'!}\hat{Q}^{A}_{I_{l}}(t)\hat{Q}^{B}_{J_{l'}}(t)\partial^{A}_{iI_{l}J_{l'}}\left(\frac{\text{e}^{-m_{\text{s}}|\boldsymbol{x}^{A}-\boldsymbol{x}^{B}|}}{|\boldsymbol{x}^{A}-\boldsymbol{x}^{B}|}\right),
\end{eqnarray}
where
\begin{eqnarray}
\label{equ3.55}&&\hat{M}^{A}_{I_{l}}(t)=\int_{A}\rho(t,\boldsymbol{x})\left(x-x^{A}\right)_{\langle I_{l}\rangle}\text{d}^{3}x,\\
\label{equ3.56}&&\hat{M}^{B}_{J_{l'}}(t)=\int_{B}\rho(t,\boldsymbol{x}')\left(x'-x^{B}\right)_{\langle J_{l'}\rangle}\text{d}^{3}x'
\end{eqnarray}
are the mass multipole moments of bodies $A$ and $B$, and
\begin{eqnarray}
\label{equ3.57}&&\hat{Q}^{A}_{I_{l}}(t)=\int_{A}\rho(t,\boldsymbol{x})\left(x-x^{A}\right)_{\langle I_{l}\rangle}\updelta_{l}\left(m_{\text{s}}|\boldsymbol{x}-\boldsymbol{x}^{A}|\right)\text{d}^{3}x,\\
\label{equ3.58}&&\hat{Q}^{B}_{J_{l'}}(t)=\int_{B}\rho(t,\boldsymbol{x}')\left(x'-x^{B}\right)_{\langle J_{l'}\rangle}\updelta_{l'}\left(m_{\text{s}}|\boldsymbol{x}'-\boldsymbol{x}^{B}|\right)\text{d}^{3}x'
\end{eqnarray}
are their scalar multipole moments. This equation yields the multipole expansion for the center-of-mass acceleration of body $A$, and it is shown that the expansion consists of the Coulomb-type part and the Yukawa-type part, which are, respectively, characterized by the multi-index derivatives of $1/|\boldsymbol{x}^{A}-\boldsymbol{x}^{B}|$ and $\text{e}^{-m_{\text{s}}|\boldsymbol{x}^{A}-\boldsymbol{x}^{B}|}/|\boldsymbol{x}^{A}-\boldsymbol{x}^{B}|$. Obviously, the Coulomb-type part is encoded by the products of the mass multipole moments of the body with those of other bodies and is exactly the same as its counterpart in GR~\cite{Eric2014}. The Yukawa-type part is encoded by the products of the scalar multipole moments of the body with those of other bodies and represents the modification to the Coulomb-type part introduced in $f(R)$ gravity. One could infer that once the mass and scalar multipole moments of the bodies are specified as functions of time, the multipole expansions of the center-of-mass accelerations of all the bodies form a complete set of equations of motion for the orbital dynamics of the system. Thus, Eq.~(\ref{equ3.54}) should be the starting point to understand the orbital motion of the system within the framework of $f(R)$ gravity.

The expressions for the mass and scalar multipole moments imply that both of them depend on the mass distribution of the source. The definitions of the mass multipole moments of bodies are identical to the those in GR, so for a body,  its mass monopole moment is the total mass, and its mass dipole moment is zero. In contrast, the scalar multipole moments of bodies lack these special properties because in their definitions, the integrals are regulated by a radial function $\updelta_{l}$, and thus, from Eq.~(\ref{equ3.40}), one can verify that the scalar monopole moment of a body
\begin{eqnarray}
\label{equ3.59}&&\int_{A}\rho(t,\boldsymbol{x})\frac{\sinh{\left(m_{\text{s}}|\boldsymbol{x}-\boldsymbol{x}^{A}|\right)}}{m_{\text{s}}|\boldsymbol{x}-\boldsymbol{x}^{A}|}\text{d}^{3}x=:Q^{A}(t)
\end{eqnarray}
is not equal to its total mass, and its scalar dipole moment does not vanish. Despite this, the scalar multipole moment, similar to the mass multipole moment, still bears an important property that all the non-monopole moments of an ideal spherical body are zero. In fact, by means of formula~(\ref{equ2.9.5}) and Eq.~(\ref{equ3.35}), one can find that the angle integrals in Eqs.~(\ref{equ3.55})--(\ref{equ3.58}) for $l\geqslant1$ vanish, which thus validates the property. Based on this property, if all the bodies in the system are ideal spheres, the center-of-mass acceleration of body $A$ only retains the monopole terms,
\begin{eqnarray}
\label{equ3.60}a_{i}^{A}&=&-\sum_{B\neq A}\frac{Gm^{B}}{|\boldsymbol{x}^{A}-\boldsymbol{x}^{B}|^{3}}\left(x^{A}-x^{B}\right)_{i}+\frac{G}{3}\sum_{B\neq A}\frac{Q^{A}}{m^{A}}Q^{B}\partial^{A}_{i}\left(\frac{\text{e}^{-m_{\text{s}}|\boldsymbol{x}^{A}-\boldsymbol{x}^{B}|}}{|\boldsymbol{x}^{A}-\boldsymbol{x}^{B}|}\right).\qquad
\end{eqnarray}
The mass monopole term corresponds to the result in GR, and if only this term is considered, Eq.~(\ref{equ3.60}) is equivalent to that governing the motion of a point body within a system composed of point sources. In $f(R)$ gravity, the situation is different due to the existence of the scalar monopole moments. When the system comprises $N$ point sources, the mass density of body $A$ is given by
\begin{eqnarray}
\label{equ3.61}&&\rho(t,\boldsymbol{x})=m^{A}\delta\left(\boldsymbol{x}-\boldsymbol{x}^{A}_{0}\right)
\end{eqnarray}
with $\boldsymbol{x}^{A}_{0}$ as its position vector. Given that the function $\updelta_{l}$ satisfies the following property~\cite{Wu:2023qjp,Arfken1985}
\begin{eqnarray}
\label{equ3.62}\lim_{z\rightarrow0}\updelta_{l}(z)=1,
\end{eqnarray}
one can prove that the scalar monopole moment of the point body $A$ is equal to its total mass
\begin{eqnarray}
\label{equ3.63}&&Q^{A}(t)=m^{A}.
\end{eqnarray}
As a result, in this case, its acceleration is further reduced to
\begin{eqnarray}
\label{equ3.64}a_{i}^{A}&=&-\sum_{B\neq A}\frac{Gm^{B}}{|\boldsymbol{x}^{A}-\boldsymbol{x}^{B}|^{3}}\left(x^{A}-x^{B}\right)_{i}+\frac{G}{3}\sum_{B\neq A}m^{B}\partial^{A}_{i}\left(\frac{\text{e}^{-m_{\text{s}}|\boldsymbol{x}^{A}-\boldsymbol{x}^{B}|}}{|\boldsymbol{x}^{A}-\boldsymbol{x}^{B}|}\right).
\end{eqnarray}
The comparison of Eqs.~(\ref{equ3.64}) and (\ref{equ3.60}) explicitly demonstrates that, in $f(R)$ gravity, for a system composed of multiple ideal spheres, the sizes of the bodies play an essential role in determining their orbital motions. This stands in contrast to GR, where a spherical body is exactly equivalent to a point source. It also differs from massless scalar-tensor theories~\cite{Kopeikin2019}, where the equivalence between a spherical body and a point source depends on the definition of the multipole moments. 
Additionally, other features of $f(R)$ gravity can also be found from Eq.~(\ref{equ3.60}). Namely, the equation of motion for a spherical body is generally time-dependent, and its own monopole moment plays an important role in defining its orbital motion.

Now, let us return to the situation of non-spherical bodies. To facilitate a convenient analysis of the roles played by the higher-order multipole moments, the multipole expansion of the center-of-mass acceleration of body $A$ is recast into a friendlier form, namely
\begin{eqnarray}
\label{equ3.65}a_{i}^{A}&=&-\sum_{B\neq A}\frac{Gm^{B}}{|\boldsymbol{x}^{A}-\boldsymbol{x}^{B}|^{3}}\left(x^{A}-x^{B}\right)_{i}\notag\\
&&+G\sum_{B\neq A}\sum_{l=2}^{\infty}\frac{1}{l!}\left[\frac{m^{B}}{m^{A}}\hat{M}^{A}_{I_{l}}+(-1)^{l}\hat{M}^{B}_{I_{l}}\right]\partial^{A}_{iI_{l}}\left(\frac{1}{|\boldsymbol{x}^{A}-\boldsymbol{x}^{B}|}\right)\notag
\end{eqnarray}
\begin{eqnarray}
&&+G\sum_{B\neq A}\sum_{l=2}^{\infty}\sum_{l'=2}^{\infty}\frac{(-1)^{l'}}{l!\,l'!}\frac{1}{m^{A}}\hat{M}^{A}_{I_{l}}\hat{M}^{B}_{J_{l'}}\partial^{A}_{iI_{l}J_{l'}}\left(\frac{1}{|\boldsymbol{x}^{A}-\boldsymbol{x}^{B}|}\right)\notag\\
&&+\frac{G}{3}\sum_{B\neq A}\frac{Q^{A}}{m^{A}}Q^{B}\partial^{A}_{i}\left(\frac{\text{e}^{-m_{\text{s}}|\boldsymbol{x}^{A}-\boldsymbol{x}^{B}|}}{|\boldsymbol{x}^{A}-\boldsymbol{x}^{B}|}\right)\notag\\
&&+\frac{G}{3}\sum_{B\neq A}\sum_{l=1}^{\infty}\frac{1}{l!}\left[\frac{Q^{B}}{m^{A}}\hat{Q}^{A}_{I_{l}}+(-1)^{l}\frac{Q^{A}}{m^{A}}\hat{Q}^{B}_{I_{l}}\right]\partial^{A}_{iI_{l}}\left(\frac{\text{e}^{-m_{\text{s}}|\boldsymbol{x}^{A}-\boldsymbol{x}^{B}|}}{|\boldsymbol{x}^{A}-\boldsymbol{x}^{B}|}\right)\notag\\
&&+\frac{G}{3}\sum_{B\neq A}\sum_{l=1}^{\infty}\sum_{l'=1}^{\infty}\frac{(-1)^{l'}}{l!\,l'!}\frac{1}{m^{A}}\hat{Q}^{A}_{I_{l}}\hat{Q}^{B}_{J_{l'}}\partial^{A}_{iI_{l}J_{l'}}\left(\frac{\text{e}^{-m_{\text{s}}|\boldsymbol{x}^{A}-\boldsymbol{x}^{B}|}}{|\boldsymbol{x}^{A}-\boldsymbol{x}^{B}|}\right),
\end{eqnarray}
where the sums in the Coulomb-type and Yukawa-type parts are both divided into terms linear in the non-monopole moments and terms involving products of the non-monopole moments. Firstly, as implied by the terms involving $\hat{M}^{B}_{I_{l}}$ and $Q^{A}\hat{Q}^{B}_{I_{l}}/m^{A}$ in this equation, the motion of body $A$ is influenced by the deformation-induced gravitational potentials generated by the other bodies. Secondly, from the terms involving $m^{B}\hat{M}^{A}_{I_{l}}/m^{A}$ and $Q^{B}\hat{Q}^{A}_{I_{l}}/m^{A}$, it can be seen that the motion of body $A$ is also influenced by its own non-spherical mass distribution coupled to the monopole potentials caused by other bodies, which suggests that the size and shape of a body can affect its orbital motion. Finally, the last lines in the Coulomb-type and Yukawa-type parts illustrate that the motion of body $A$ is influenced by the interactions between its own higher-order multipole moments and those of other bodies. In GR, the leading effect of this kind comes from a quadrupole-quadrupole coupling. It should be noted that the emergence of a leading dipole-dipole interaction, as well as the overall structure of finite-size effects captured by multipole moments, is broadly consistent with expectations from an EFT treatment of extended bodies in massless scalar-tensor theories \cite{Kopeikin2019}. Therefore, the presence of dipole effects is not unique to $f(R)$ gravity.

To conclude the discussion of Eq.~(\ref{equ3.65}), the distinction regarding the approximation scheme needs to be highlighted. In GR, only the Coulomb-type part is kept on the right side of Eq.~(\ref{equ3.65}), and since $$\left(x-x^{A}\right)_{\langle I_{l}\rangle}\left(x'-x^{B}\right)_{\langle J_{l'}\rangle}\partial^{A}_{iI_{l}J_{l'}}\left(\frac{1}{|\boldsymbol{x}^{A}-\boldsymbol{x}^{B}|}\right)$$ in Eq.~(\ref{equ3.51}) scales as
\begin{eqnarray}
\label{equ3.66}\frac{1}{|\boldsymbol{x}^{A}-\boldsymbol{x}^{B}|^{2}}\left(\frac{|\boldsymbol{x}-\boldsymbol{x}^{A}|}{|\boldsymbol{x}^{A}-\boldsymbol{x}^{B}|}\right)^{l}\left(\frac{|\boldsymbol{x}'-\boldsymbol{x}^{B}|}{|\boldsymbol{x}^{A}-\boldsymbol{x}^{B}|}\right)^{l'}
\end{eqnarray}
and the typical size of each body is far less than the typical separation between bodies, the mass multipole terms will get progressively smaller as the orders increase. Therefore, in this case, the corresponding equation can serve as a starting point of an approximation scheme for the system, and thus, the sums can be truncated according to the required precision. However, this conclusion does not apply to the result given by Eq.~(\ref{equ3.65}) in $f(R)$ gravity. For body $A$, based on Eq.~(\ref{equ3.35}) and the relationship~(\ref{equ3.41}) between $\updelta_{l}(z)$ and $\text{i}_{l}(z)$, we have
\begin{eqnarray}
\label{equ3.67}\updelta_{l}\left(m_{\text{s}}|\boldsymbol{x}-\boldsymbol{x}^{A}|\right)\left(x-x^{A}\right)_{\langle I_{l}\rangle}&=&\frac{1}{m_{\text{s}}^{l}}\updelta_{l}\left(m_{\text{s}}r\right)(m_{\text{s}}r)^l\hat{N}_{I_{l}}(\theta,\varphi)\notag\\
&=&\frac{(2l+1)!!}{m_{\text{s}}^{l}}\text{i}_{l}\left(m_{\text{s}}|\boldsymbol{x}-\boldsymbol{x}^{A}|\right)\hat{N}_{I_{l}}(\theta,\varphi).
\end{eqnarray}
In order to ensure that the gravitational potential of $f(R)$ gravity presented in Eqs.~(\ref{equ3.2})--(\ref{equ3.4}) satisfies Solar-system observational constraints, the parameter $m_{\text{s}}$  must be taken sufficiently large.
Starting from this condition, by use of  the asymptotic expansion of $\text{i}_{l}(z)$ given in Eq.~(\ref{equ3.39}),
the leading-order term of $\updelta_{l}\left(m_{\text{s}}|\boldsymbol{x}-\boldsymbol{x}^{A}|\right)\left(x-x^{A}\right)_{\langle I_{l}\rangle}$ is
\begin{eqnarray}
\label{equ3.68}\frac{1}{m_{\text{s}}^{l}}\frac{\text{e}^{m_{\text{s}}|\boldsymbol{x}-\boldsymbol{x}^{A}|}}{2m_{\text{s}}|\boldsymbol{x}-\boldsymbol{x}^{A}|}.
\end{eqnarray}
Similarly, the leading-order term of $\updelta_{l'}\left(m_{\text{s}}|\boldsymbol{x}'-\boldsymbol{x}^{B}|\right)\left(x'-x^{B}\right)_{\langle J_{l'}\rangle}$ for body $B$ is
\begin{eqnarray}
\label{equ3.69}\frac{1}{m_{\text{s}}^{l'}}\frac{\text{e}^{m_{\text{s}}|\boldsymbol{x}'-\boldsymbol{x}^{B}|}}{2m_{\text{s}}|\boldsymbol{x}'-\boldsymbol{x}^{B}|}.
\end{eqnarray}
Besides these, the condition that $m_{\text{s}}$ is sufficiently large also implies that the leading-order term of the multi-index derivatives of $\text{e}^{-m_{\text{s}}|\boldsymbol{x}^{A}-\boldsymbol{x}^{B}|}/|\boldsymbol{x}^{A}-\boldsymbol{x}^{B}|$ in Eq.~(\ref{equ3.52}) should be
\begin{eqnarray}
\label{equ3.70}&&m_{\text{s}}^{l+l'+1}\frac{\text{e}^{-m_{\text{s}}|\boldsymbol{x}^{A}-\boldsymbol{x}^{B}|}}{|\boldsymbol{x}^{A}-\boldsymbol{x}^{B}|}.
\end{eqnarray}
Combining these three results, the leading-order contribution of
$$\updelta_{l}\left(m_{\text{s}}|\boldsymbol{x}-\boldsymbol{x}^{A}|\right)\left(x-x^{A}\right)_{\langle I_{l}\rangle}\updelta_{l'}\left(m_{\text{s}}|\boldsymbol{x}'-\boldsymbol{x}^{B}|\right)\left(x'-x^{B}\right)_{\langle J_{l'}\rangle}\partial^{A}_{iI_{l}J_{l'}}\left(\frac{\text{e}^{-m_{\text{s}}|\boldsymbol{x}^{A}-\boldsymbol{x}^{B}|}}{|\boldsymbol{x}^{A}-\boldsymbol{x}^{B}|}\right)$$
in Eq.~(\ref{equ3.52}) reads
\begin{eqnarray}
\label{equ3.71}&&m_{\text{s}}\frac{\text{e}^{m_{\text{s}}|\boldsymbol{x}-\boldsymbol{x}^{A}|}}{2m_{\text{s}}|\boldsymbol{x}-\boldsymbol{x}^{A}|}\frac{\text{e}^{m_{\text{s}}|\boldsymbol{x}'-\boldsymbol{x}^{B}|}}{2m_{\text{s}}|\boldsymbol{x}'-\boldsymbol{x}^{B}|}
\frac{\text{e}^{-m_{\text{s}}|\boldsymbol{x}^{A}-\boldsymbol{x}^{B}|}}{|\boldsymbol{x}^{A}-\boldsymbol{x}^{B}|},
\end{eqnarray}
which is independent of the multipole orders $l$ or $l'$. As a consequence, it cannot be concluded that the scalar multipole terms in Eq.~(\ref{equ3.65}) decay systematically with increasing order. This means that, in contrast to the GR case, the sums appearing in the Yukawa-type part cannot be arbitrarily truncated in practical calculations.
\section{Conserved quantities and spin dynamics in the isolated self-gravitating system~\label{Sec:fourth}}
\subsection{Conserved quantities of the system}
According to the initial assumption in the introduction, each body within the isolated self-gravitating system is composed of a perfect fluid, which implies that all the bodies form a perfect fluid system.
Similar to the case in GR,  starting from the Euler's equation~(\ref{equ3.5}), many important global conservation laws can be established in $f(R)$ gravity at the Newtonian order, and from them, the conserved quantities such as the total momentum, angular momentum, and energy can also be provided. In this subsection, we aim to present the expressions for these conserved quantities in the isolated self-gravitating system.

By the definition in classical mechanics, the total momentum of the entire system is given by
\begin{eqnarray}
\label{equ4.1}&&\boldsymbol{P}:=\sum_{A=1}^{N}\int_{A}\rho(t,\boldsymbol{x})\boldsymbol{v}(t,\boldsymbol{x})\,\text{d}^{3}x=\sum_{A=1}^{N}m^{A}\boldsymbol{v}^{A},
\end{eqnarray}
where the expression for the center-of-mass velocity of a body is used. As expected, this suggests that the total momentum of the system is the sum of each body's center-of-mass momentum. The conservation of the total momentum of the system can be verified by considering
\begin{eqnarray}
\label{equ4.2}\frac{\text{d}\boldsymbol{P}}{\text{d}t}&=&\sum_{A=1}^{N}\int_{A}\rho\frac{\text{d}\boldsymbol{v}}{\text{d}t}\,\text{d}^{3}x=\sum_{A=1}^{N}\int_{A}\rho\nabla\varPhi\,\text{d}^{3}x-\sum_{A=1}^{N}\int_{A}\nabla p\,\text{d}^{3}x\notag\\
&=&G\sum_{A=1}^{N}\sum_{B=1}^{N}\int_{A}\int_{B}\rho\rho'\nabla\left(\frac{1}{|\boldsymbol{x}-\boldsymbol{x}'|}\right)\,\text{d}^{3}x\text{d}^{3}x'\notag\\
&&+\frac{G}{3}\sum_{A=1}^{N}\sum_{B=1}^{N}\int_{A}\int_{B}\rho\rho'\nabla\left(\frac{\text{e}^{-m_{\text{s}}|\boldsymbol{x}-\boldsymbol{x}'|}}{|\boldsymbol{x}-\boldsymbol{x}'|}\right)\,\text{d}^{3}x\text{d}^{3}x'=0.
\end{eqnarray}
In the first and second steps, the formula~(\ref{equ3.12}) and the Euler's equation~(\ref{equ3.5}) are applied, whereas in the third step, Eqs.~(\ref{equ3.14}) and~(\ref{equ3.1})--(\ref{equ3.4}) are employed. As to the last step, it is valid due to
the following two identities,
\begin{eqnarray}
\label{equ4.3}&&\sum_{A=1}^{N}\sum_{B=1}^{N}\int_{A}\int_{B}\rho\rho'\frac{\partial}{\partial x_{i}}\left(\frac{1}{|\boldsymbol{x}-\boldsymbol{x}'|}\right)\,\text{d}^{3}x\text{d}^{3}x'\notag\\
&=&\sum_{B=1}^{N}\sum_{A=1}^{N}\int_{B}\int_{A}\rho'\rho\frac{\partial}{\partial x'_{i}}\left(\frac{1}{|\boldsymbol{x}'-\boldsymbol{x}|}\right)\,\text{d}^{3}x'\text{d}^{3}x\notag\\
&=&-\sum_{A=1}^{N}\sum_{B=1}^{N}\int_{A}\int_{B}\rho\rho'\frac{\partial}{\partial x_{i}}\left(\frac{1}{|\boldsymbol{x}-\boldsymbol{x}'|}\right)\,\text{d}^{3}x\text{d}^{3}x'
\end{eqnarray}
and
\begin{eqnarray}
&&\sum_{A=1}^{N}\sum_{B=1}^{N}\int_{A}\int_{B}\rho\rho'\frac{\partial}{\partial x_{i}}\left(\frac{\text{e}^{-m_{\text{s}}|\boldsymbol{x}-\boldsymbol{x}'|}}{|\boldsymbol{x}-\boldsymbol{x}'|}\right)\,\text{d}^{3}x\text{d}^{3}x'\notag\\
&=&\sum_{B=1}^{N}\sum_{A=1}^{N}\int_{B}\int_{A}\rho'\rho\frac{\partial}{\partial x'_{i}}\left(\frac{\text{e}^{-m_{\text{s}}|\boldsymbol{x}'-\boldsymbol{x}|}}{|\boldsymbol{x}'-\boldsymbol{x}|}\right)\,\text{d}^{3}x'\,\text{d}^{3}x\notag\\
\label{equ4.4}
&=&-\sum_{A=1}^{N}\sum_{B=1}^{N}\int_{A}\int_{B}\rho\rho'\frac{\partial}{\partial x_{i}}\left(\frac{\text{e}^{-m_{\text{s}}|\boldsymbol{x}-\boldsymbol{x}'|}}{|\boldsymbol{x}-\boldsymbol{x}'|}\right)\,\text{d}^{3}x\,\text{d}^{3}x',
\end{eqnarray}
and the derivation processes of them are similar to those of Eqs.~(\ref{equ3.18}) and (\ref{equ3.19}).  Obviously, even though the gravitational potential produced by all the bodies in the system under $f(R)$ gravity differs from that in GR, since the function $\text{e}^{-m_{\text{s}}|\boldsymbol{x}-\boldsymbol{x}'|}/|\boldsymbol{x}-\boldsymbol{x}'|$ also depends solely on the difference between $\boldsymbol{x}$ and $\boldsymbol{x}'$ the conservation of the total momentum of the system is still maintained. As a result, with the help of Eq.~(\ref{equ3.11}), we can further obtain
\begin{eqnarray}
\label{equ4.5}&&\sum_{A=1}^{N}m^{A}\boldsymbol{a}^{A}=0,
\end{eqnarray}
which actually implies that Newton's third law still holds in $f(R)$ gravity at the Newtonian order. In addition, the conservation of the total momentum also means that the barycenter of the system, defined by
\begin{eqnarray}
\label{equ4.6}&&\boldsymbol{X}:=\frac{1}{m}\sum_{A=1}^{N}m^{A}\boldsymbol{x}^{A}\quad\text{with}\quad m:=\sum_{A=1}^{N}m^{A},
\end{eqnarray}
always moves at a constant velocity, which is exactly the same as the situation in classical celestial mechanics. It will be seen that the above property of the function $\text{e}^{-m_{\text{s}}|\boldsymbol{x}-\boldsymbol{x}'|}/|\boldsymbol{x}-\boldsymbol{x}'|$ also ensures the conservation of the total angular momentum of the entire system, and we will demonstrate this conclusion in what follows. By the conventional definition, the total angular momentum of the system is defined by
\begin{eqnarray}
\label{equ4.7}&&\boldsymbol{J}:=\sum_{A=1}^{N}\int_{A}\rho\boldsymbol{x}\times\boldsymbol{v}\,\text{d}^{3}x=\sum_{A=1}^{N}\left(\boldsymbol{L}^{A}+\boldsymbol{S}^{A}\right)
\end{eqnarray}
with
\begin{eqnarray}
\label{equ4.8}&&\boldsymbol{L}^{A}:=m^{A}\boldsymbol{x}^{A}\times\boldsymbol{v}^{A},\\
\label{equ4.9}&&\boldsymbol{S}^{A}:=\int_{A}\rho\left(\boldsymbol{x}-\boldsymbol{x}^{A}\right)\times\left(\boldsymbol{v}-\boldsymbol{v}^{A}\right)\,\text{d}^{3}x
\end{eqnarray}
as the orbital and spin angular momenta of body $A$, respectively, and their expressions can directly be obtained by means of the definitions for the center-of-mass velocity and acceleration of a body. In order to demonstrate that $\boldsymbol{J}$ is a conserved quantity of the system, we need to write
\begin{eqnarray}
\label{equ4.10}\frac{\text{d}\boldsymbol{J}}{\text{d}t}&=&\sum_{A=1}^{N}\int_{A}\rho\boldsymbol{x}\times\frac{\text{d}\boldsymbol{v}}{\text{d}t}\,\text{d}^{3}x\notag\\
&=&\sum_{A=1}^{N}\int_{A}\rho\boldsymbol{x}\times\nabla\varPhi\,\text{d}^{3}x+\sum_{A=1}^{N}\int_{A}\nabla\times\left(p\boldsymbol{x}\right)\,\text{d}^{3}x\notag\\
&=&G\sum_{A=1}^{N}\sum_{B=1}^{N}\int_{A}\int_{B}\rho\rho'\boldsymbol{x}\times\nabla\left(\frac{1}{|\boldsymbol{x}-\boldsymbol{x}'|}\right)\,\text{d}^{3}x\text{d}^{3}x'\notag\\
&&+\frac{G}{3}\sum_{A=1}^{N}\sum_{B=1}^{N}\int_{A}\int_{B}\rho\rho'\boldsymbol{x}\times\nabla\left(\frac{\text{e}^{-m_{\text{s}}|\boldsymbol{x}-\boldsymbol{x}'|}}{|\boldsymbol{x}-\boldsymbol{x}'|}\right)\,\text{d}^{3}x\text{d}^{3}x',
\end{eqnarray}
where the derivation of each step follows analogously to that in Eq.~(\ref{equ4.2}), and especially, it ought to be noted that in the third step,
\begin{eqnarray}
\label{equ4.11}&&\int_{A}\nabla\times\left(p\boldsymbol{x}\right)\,\text{d}^{3}x=\oint_{\partial A}\text{d}\boldsymbol{S}\times\left(p\boldsymbol{x}\right)=0
\end{eqnarray}
is invoked due to the vanishing pressure on the boundary $\partial A$ of the integration region. Again, based on the aforementioned property that both $1/|\boldsymbol{x}-\boldsymbol{x}'|$ and $\text{e}^{-m_{\text{s}}|\boldsymbol{x}-\boldsymbol{x}'|}/|\boldsymbol{x}-\boldsymbol{x}'|$ depend solely on $\boldsymbol{x}-\boldsymbol{x}'$, the two integral terms in Eq.~(\ref{equ4.10}) are evaluated as
\begin{eqnarray}
\label{equ4.12}&&\sum_{A=1}^{N}\sum_{B=1}^{N}\int_{A}\int_{B}\rho\rho'\epsilon_{ijk}x_{j}\frac{\partial}{\partial x_{k}}\left(\frac{1}{|\boldsymbol{x}-\boldsymbol{x}'|}\right)\,\text{d}^{3}x\text{d}^{3}x'\notag\\
&=&\frac{1}{2}\sum_{A=1}^{N}\sum_{B=1}^{N}\int_{A}\int_{B}\rho\rho'\epsilon_{ijk}(x-x')_{j}\frac{\partial}{\partial x_{k}}\left(\frac{1}{|\boldsymbol{x}-\boldsymbol{x}'|}\right)\,\text{d}^{3}x\text{d}^{3}x'\notag\\
&=&\frac{1}{2}\sum_{A=1}^{N}\sum_{B=1}^{N}\int_{A}\int_{B}\rho\rho'\frac{\epsilon_{ijk}(x-x')_{j}(x-x')_{k}}{|\boldsymbol{x}-\boldsymbol{x}'|}\frac{\text{d}}{\text{d}|\boldsymbol{x}-\boldsymbol{x}'|}\left(\frac{1}{|\boldsymbol{x}-\boldsymbol{x}'|}\right)\,\text{d}^{3}x\text{d}^{3}x'\notag\\
&=&0
\end{eqnarray}
and
\begin{eqnarray}
\label{equ4.13}&&\sum_{A=1}^{N}\sum_{B=1}^{N}\int_{A}\int_{B}\rho\rho'\epsilon_{ijk}x_{j}\frac{\partial}{\partial x_{k}}\left(\frac{\text{e}^{-m_{\text{s}}|\boldsymbol{x}-\boldsymbol{x}'|}}{|\boldsymbol{x}-\boldsymbol{x}'|}\right)\,\text{d}^{3}x\text{d}^{3}x'\notag\\
&=&\frac{1}{2}\sum_{A=1}^{N}\sum_{B=1}^{N}\int_{A}\int_{B}\rho\rho'\epsilon_{ijk}(x-x')_{j}\frac{\partial}{\partial x_{k}}\left(\frac{\text{e}^{-m_{\text{s}}|\boldsymbol{x}-\boldsymbol{x}'|}}{|\boldsymbol{x}-\boldsymbol{x}'|}\right)\,\text{d}^{3}x\text{d}^{3}x'\notag\\
&=&\frac{1}{2}\sum_{A=1}^{N}\sum_{B=1}^{N}\int_{A}\int_{B}\rho\rho'\frac{\epsilon_{ijk}(x-x')_{j}(x-x')_{k}}{|\boldsymbol{x}-\boldsymbol{x}'|}\frac{\text{d}}{\text{d}|\boldsymbol{x}-\boldsymbol{x}'|}\left(\frac{\text{e}^{-m_{\text{s}}|\boldsymbol{x}-\boldsymbol{x}'|}}{|\boldsymbol{x}-\boldsymbol{x}'|}\right)\,\text{d}^{3}x\text{d}^{3}x'\notag\\
&=&0,
\end{eqnarray}
and thus, we arrive at the conclusion that $\text{d}\boldsymbol{J}/\text{d}t=0$. Equation~(\ref{equ4.7}) shows that the total angular momentum of the system is the sum of the orbital and spin angular momenta for all bodies. However, while the total angular momentum is conserved, the orbital or spin angular momentum of each body is not generally conserved. Normally, the spin angular momentum of a body within the system will be influenced by the motions of other bodies, so in practical applications, establishing the equation of motion for the spin angular momentum of each body becomes a critical task. In the next subsection, we will address this issue in detail.

As for the proof of total energy conservation in the system, it is a bit more involved. As per fluid mechanics, the total energy $E$ of the entire system is composed of the kinetic energy
\begin{eqnarray}
\label{equ4.14}&&T(t):=\sum_{A=1}^{N}\frac{1}{2}\int_{A}\rho\boldsymbol{v}^{2}\,\text{d}^{3}x,
\end{eqnarray}
the gravitational potential energy
\begin{eqnarray}
\label{equ4.15}&&\varOmega(t):=-\sum_{A=1}^{N}\frac{1}{2}\int_{A}\rho\varPhi\,\text{d}^{3}x,
\end{eqnarray}
and the internal thermodynamic energy
\begin{eqnarray}
\label{equ4.16}&&E_{\text{int}}(t):=\sum_{A=1}^{N}\int_{A}\rho\varPi\,\text{d}^{3}x,
\end{eqnarray}
where $\varPi(t,\boldsymbol{x})$ is the specific internal energy. The definitions of the above three types of energies are clear. The kinetic energy of the system is defined as the sum of the kinetic energies of all fluid elements within all bodies. In view of the fact that $\rho(t,\boldsymbol{x})\varPi(t,\boldsymbol{x})$ is the density of the internal thermodynamic energy, the internal thermodynamic energy of the system is also the sum of the internal energies of all fluid elements within all bodies. By inserting the expression for the gravitational potential given in Eqs.~(\ref{equ3.1}) and (\ref{equ3.2}), the gravitational potential energy of the system can be written as
\begin{eqnarray}
\label{equ4.17}\varOmega
&=&-\sum_{A=1}^{N}\sum_{B=1}^{N}\frac{1}{2}\int_{A}\rho U^{B}\,\text{d}^{3}x-\sum_{A=1}^{N}\sum_{B=1}^{N}\frac{1}{2}\int_{A}\rho Y^{B}\,\text{d}^{3}x,
\end{eqnarray}
\noindent which is distinct from the result in the classical celestial mechanics because of the presence of the second term. Within the framework of Newtonian $f(R)$ gravity, the validity of this definition for the gravitational potential energy of the system needs to be ultimately confirmed by the conservation of the total energy, and now, let us proceed to the topic. From the above discussions, the total energy of the system should be
\begin{eqnarray}
\label{equ4.18}&&E:=T(t)+\varOmega(t)+E_{\text{int}}(t).
\end{eqnarray}
We begin with the kinetic energy $T$. By employing the formula~(\ref{equ3.12}) and the Euler's equation~(\ref{equ3.5}), its time derivative is
\begin{eqnarray}
\label{equ4.19}\frac{\text{d}T}{\text{d}t}&=&\sum_{A=1}^{N}\int_{A}\rho\boldsymbol{v}\cdot\frac{\text{d}\boldsymbol{v}}{\text{d}t}\,\text{d}^{3}x=\sum_{A=1}^{N}\int_{A}\rho\boldsymbol{v}\cdot\nabla\varPhi\,\text{d}^{3}x+\sum_{A=1}^{N}\int_{A}p\nabla\cdot\boldsymbol{v}\,\text{d}^{3}x,
\end{eqnarray}
where based on Gauss's theorem and $p=0$ on the boundary $\partial A$ of the integration region, the identity
\begin{eqnarray}
\label{equ4.20}&&\int_{A}\nabla\cdot(p\boldsymbol{v})\,\text{d}^{3}x=\oint_{\partial A}p\boldsymbol{v}\cdot\text{d}\boldsymbol{S}=0
\end{eqnarray}
is invoked in the last step.  After plugging Eqs.~(\ref{equ3.1})--(\ref{equ3.4}) and performing the variable swapping ($A\leftrightarrow B$ and $\boldsymbol{x}\leftrightarrow \boldsymbol{x}'$), the first term in Eq.~(\ref{equ4.19}) can be explicitly expressed as
\begin{eqnarray}
\label{equ4.21}\sum_{A=1}^{N}\int_{A}\rho\boldsymbol{v}\cdot\nabla\varPhi\,\text{d}^{3}x&=&G\sum_{A=1}^{N}\sum_{B=1}^{N}\int_{A}\int_{B}\rho\rho'\boldsymbol{v}\cdot\nabla\left(\frac{1}{|\boldsymbol{x}-\boldsymbol{x}'|}\right)\,\text{d}^{3}x\text{d}^{3}x'\notag\\
&&+\frac{G}{3}\sum_{A=1}^{N}\sum_{B=1}^{N}\int_{A}\int_{B}\rho\rho'\boldsymbol{v}\cdot\nabla\left(\frac{\text{e}^{-m_{\text{s}}|\boldsymbol{x}-\boldsymbol{x}'|}}{|\boldsymbol{x}-\boldsymbol{x}'|}\right)\,\text{d}^{3}x\text{d}^{3}x'\notag\\
&=&G\sum_{B=1}^{N}\sum_{A=1}^{N}\int_{B}\int_{A}\rho'\rho\boldsymbol{v}'\cdot\nabla'\left(\frac{1}{|\boldsymbol{x}'-\boldsymbol{x}|}\right)\,\text{d}^{3}x'\text{d}^{3}x\notag\\
&&+\frac{G}{3}\sum_{B=1}^{N}\sum_{A=1}^{N}\int_{B}\int_{A}\rho'\rho\boldsymbol{v}'\cdot\nabla'\left(\frac{\text{e}^{-m_{\text{s}}|\boldsymbol{x}'-\boldsymbol{x}|}}{|\boldsymbol{x}'-\boldsymbol{x}|}\right)\,\text{d}^{3}x'\text{d}^{3}x,\qquad
\end{eqnarray}
in which, $\boldsymbol{v}':=\boldsymbol{v}(t,\boldsymbol{x}')$ and $\nabla'$ is the gradient operator associated with $\boldsymbol{x}'$. With this result, the first term in Eq.~(\ref{equ4.19}) can be rewritten as follows:
\begin{eqnarray}
\sum_{A=1}^{N}\int_{A}\rho\boldsymbol{v}\cdot\nabla\varPhi\,\text{d}^{3}x&=&\frac{G}{2}\sum_{A=1}^{N}\sum_{B=1}^{N}\int_{A}\int_{B}\rho\rho'\left(\boldsymbol{v}\cdot\nabla+\boldsymbol{v}'\cdot\nabla'\right)\left(\frac{1}{|\boldsymbol{x}-\boldsymbol{x}'|}\right)\,\text{d}^{3}x\text{d}^{3}x'\notag\\
\label{equ4.22}&&+\frac{G}{6}\sum_{A=1}^{N}\sum_{B=1}^{N}\int_{A}\int_{B}\rho\rho'\left(\boldsymbol{v}\cdot\nabla+\boldsymbol{v}'\cdot\nabla'\right)\left(\frac{\text{e}^{-m_{\text{s}}|\boldsymbol{x}-\boldsymbol{x}'|}}{|\boldsymbol{x}-\boldsymbol{x}'|}\right)\,\text{d}^{3}x\text{d}^{3}x'.\quad\qquad
\end{eqnarray}
In the spirit of the convective time derivative (cf.~Eq.~(\ref{equ3.7})) in fluid mechanics, there should be
\begin{eqnarray}
\label{equ4.23}&&\left(\boldsymbol{v}\cdot\nabla+\boldsymbol{v}'\cdot\nabla'\right)\left(\frac{1}{|\boldsymbol{x}-\boldsymbol{x}'|}\right)=\frac{\text{d}}{\text{d}t}\left(\frac{1}{|\boldsymbol{x}-\boldsymbol{x}'|}\right),\\
\label{equ4.24}&&\left(\boldsymbol{v}\cdot\nabla+\boldsymbol{v}'\cdot\nabla'\right)\left(\frac{\text{e}^{-m_{\text{s}}|\boldsymbol{x}-\boldsymbol{x}'|}}{|\boldsymbol{x}-\boldsymbol{x}'|}\right)=\frac{\text{d}}{\text{d}t}\left(\frac{\text{e}^{-m_{\text{s}}|\boldsymbol{x}-\boldsymbol{x}'|}}{|\boldsymbol{x}-\boldsymbol{x}'|}\right),
\end{eqnarray}
which results in that the first term in Eq.~(\ref{equ4.19}) is finally recast into
\begin{eqnarray}
\label{equ4.25}\sum_{A=1}^{N}\int_{A}\rho\boldsymbol{v}\cdot\nabla\varPhi\,\text{d}^{3}x&=&\frac{G}{2}\sum_{A=1}^{N}\sum_{B=1}^{N}\int_{A}\int_{B}\rho\rho'\frac{\text{d}}{\text{d}t}\left(\frac{1}{|\boldsymbol{x}-\boldsymbol{x}'|}\right)\,\text{d}^{3}x\text{d}^{3}x'\notag\\
&&+\frac{G}{6}\sum_{A=1}^{N}\sum_{B=1}^{N}\int_{A}\int_{B}\rho\rho'\frac{\text{d}}{\text{d}t}\left(\frac{\text{e}^{-m_{\text{s}}|\boldsymbol{x}-\boldsymbol{x}'|}}{|\boldsymbol{x}-\boldsymbol{x}'|}\right)\,\text{d}^{3}x\text{d}^{3}x'.
\end{eqnarray}
The handling of the second term in Eq.~(\ref{equ4.19}) is simple, and the continuity equation~(\ref{equ3.6}) can give us
\begin{eqnarray}
\label{equ4.26}\sum_{A=1}^{N}\int_{A}p\nabla\cdot\boldsymbol{v}\,\text{d}^{3}x=-\sum_{A=1}^{N}\int_{A}\frac{p}{\rho}\frac{\text{d}\rho}{\text{d}t}\,\text{d}^{3}x.
\end{eqnarray}
The calculations of the time derivatives of the gravitational potential energy $\varOmega$ and the internal thermodynamic energy $E_{\text{int}}$ are direct. By applying formula~(\ref{equ3.12}) to Eq.~(\ref{equ4.17}) and substituting the expressions for the Coulomb-like and Yukawa-like potentials, we get
\begin{eqnarray}
\label{equ4.27}\frac{\text{d}\varOmega}{\text{d}t}
&=&-\frac{G}{2}\sum_{A=1}^{N}\sum_{B=1}^{N}\int_{A}\int_{B}\rho\rho'\frac{\text{d}}{\text{d}t}\left(\frac{1}{|\boldsymbol{x}-\boldsymbol{x}'|}\right)\,\text{d}^{3}x\text{d}^{3}x'\notag\\
&&-\frac{G}{6}\sum_{A=1}^{N}\sum_{B=1}^{N}\int_{A}\int_{B}\rho\rho'\frac{\text{d}}{\text{d}t}\left(\frac{\text{e}^{-m_{\text{s}}|\boldsymbol{x}-\boldsymbol{x}'|}}{|\boldsymbol{x}-\boldsymbol{x}'|}\right)\,\text{d}^{3}x\text{d}^{3}x'.
\end{eqnarray}
In Ref.~\cite{Eric2014}, for a fluid system in thermodynamic equilibrium, the first law of thermodynamics is cast in the following form
\begin{eqnarray}
\label{equ4.28}\text{d}\varPi=\frac{p}{\rho^2}\text{d}\rho,
\end{eqnarray}
and then, if it is assumed that all the bodies in the system are at all times in thermal equilibrium, by invoking formula~(\ref{equ3.12}) again, the time derivative of the internal thermodynamic energy is obtained,
\begin{eqnarray}
\label{equ4.29}\frac{\text{d}E_{\text{int}}}{\text{d}t}=\sum_{A=1}^{N}\int_{A}\frac{p}{\rho}\frac{\text{d}\rho}{\text{d}t}\,\text{d}^{3}x.
\end{eqnarray}
Combining Eqs.~(\ref{equ4.19}), (\ref{equ4.25}), (\ref{equ4.26}), (\ref{equ4.27}), and (\ref{equ4.29}), the conclusion $\text{d}E/\text{d}t=0$ is derived, which signifies that the total energy of the system in $f(R)$ gravity (at the Newtonian order) is indeed conserved.

For the isolated self-gravitating system, it is significant to provide the reasonable expressions for the kinetic energy, the gravitational potential energy, and the internal thermodynamic energy. The internal kinetic energy of a body in the system
should be defined as the kinetic energy in its center-of-mass frame, namely,
\begin{eqnarray}
\label{equ4.30}&&T^{A}:=\frac{1}{2}\int_{A}\rho\left(\boldsymbol{v}-\boldsymbol{v}^{A}\right)^{2}\,\text{d}^{3}x,
\end{eqnarray}
and then, a simple calculation reveals that the total kinetic energy of the system can be decomposed into
\begin{eqnarray}
\label{equ4.31}&&T(t)=\sum_{A=1}^{N}\left[\frac{1}{2}m^{A}\left(\boldsymbol{v}^{A}\right)^{2}+T^{A}\right].
\end{eqnarray}
Furthermore, the internal gravitational potential energy of a body is given by
\begin{eqnarray}
\label{equ4.32}&&\varOmega^{A}:=-\frac{1}{2}\int_{A}\rho\varPhi^{A}\,\text{d}^{3}x
\end{eqnarray}
and it represents the potential energy due to the gravitational interactions between fluid elements within the body in $f(R)$ gravity. Then, on the basis of the decomposition~(\ref{equ3.15}), the total gravitational potential energy of the system should be written as
\begin{eqnarray}
\label{equ4.33}&&\varOmega=\sum_{A=1}^{N}\left(\varOmega^{A}-\frac{1}{2}\int_{A}\rho\varPhi^{-A}\,\text{d}^{3}x\right),
\end{eqnarray}
where after inserting Eqs.~(\ref{equ3.21})--(\ref{equ3.23}), the second term within the bracket, being the gravitational potential energy between body $A$ and all other bodies, is expressed as
\begin{eqnarray}
\label{equ4.34}\sum_{A=1}^{N}\left(-\frac{1}{2}\int_{A}\rho\varPhi^{-A}\right)&=&-\frac{G}{2}\sum_{A=1}^{N}\sum_{B\neq A}\int_{A}\int_{B}\rho\rho'\frac{1}{|\boldsymbol{x}-\boldsymbol{x}'|}\,\text{d}^{3}x\text{d}^{3}x'\notag\\
&&-\frac{G}{6}\sum_{A=1}^{N}\sum_{B\neq A}\int_{A}\int_{B}\rho\rho'\frac{\text{e}^{-m_{\text{s}}|\boldsymbol{x}-\boldsymbol{x}'|}}{|\boldsymbol{x}-\boldsymbol{x}'|}\,\text{d}^{3}x\text{d}^{3}x'.
\end{eqnarray}
Clearly, the mass distributions of body $A$ and all other bodies play significant roles in the above gravitational potential energy, so
performing a multipole expansion for this potential energy is crucial. Given that the multipole expansions of $1/|\boldsymbol{x}-\boldsymbol{x}'|$ and $\text{e}^{-m_{\text{s}}|\boldsymbol{x}-\boldsymbol{x}'|}/|\boldsymbol{x}-\boldsymbol{x}'|$ have been provided in Eqs.~(\ref{equ3.47}) and (\ref{equ3.48}),  there is
\begin{eqnarray}
\label{equ4.35}\sum_{A=1}^{N}\left(-\frac{1}{2}\int_{A}\rho\varPhi^{-A}\right)
&=&-\frac{G}{2}\sum_{A=1}^{N}\sum_{B\neq A}\sum_{l=0}^{\infty}\sum_{l'=0}^{\infty}\frac{(-1)^{l'}}{l!\,l'!}\hat{M}^{A}_{I_{l}}\hat{M}^{B}_{J_{l'}}\partial^{A}_{I_{l}J_{l'}}\left(\frac{1}{|\boldsymbol{x}^{A}-\boldsymbol{x}^{B}|}\right)\notag\\
&&-\frac{G}{6}\sum_{A=1}^{N}\sum_{B\neq A}\sum_{l=0}^{\infty}\sum_{l'=0}^{\infty}\frac{(-1)^{l'}}{l!\,l'!}\hat{Q}^{A}_{I_{l}}\hat{Q}^{B}_{J_{l'}}\partial^{A}_{I_{l}J_{l'}}\left(\frac{\text{e}^{-m_{\text{s}}|\boldsymbol{x}^{A}-\boldsymbol{x}^{B}|}}{|\boldsymbol{x}^{A}-\boldsymbol{x}^{B}|}\right).\qquad\quad
\end{eqnarray}
By definition, the internal thermodynamic energy of a body is naturally given by
\begin{eqnarray}
\label{equ4.36}&&E^{A}_{\text{int}}:=\int_{A}\rho\varPi\,\text{d}^{3}x,
\end{eqnarray}
and thus, the internal thermodynamic energy of the system is therefore the sum of those of all the bodies,
\begin{eqnarray}
\label{equ4.37}&&E_{\text{int}}=\sum_{A=1}^{N}E^{A}_{\text{int}}.
\end{eqnarray}
With the internal kinetic energy, gravitational potential energy, and thermodynamic energy of a body, its self-energy ought to be furnished by
\begin{eqnarray}
\label{equ4.38}&&E^{A}:=T^{A}+\varOmega^{A}+E^{A}_{\text{int}},
\end{eqnarray}
and based on the previous results, we obtain the final expression for the total energy of the system,
\begin{eqnarray}
\label{equ4.39}E&=&\sum_{A=1}^{N}E^{A}+\sum_{A=1}^{N}\frac{1}{2}m^{A}\left(\boldsymbol{v}^{A}\right)^{2}\notag\\
&&-\frac{G}{2}\sum_{A=1}^{N}\sum_{B\neq A}\sum_{l=0}^{\infty}\sum_{l'=0}^{\infty}\frac{(-1)^{l'}}{l!\,l'!}\hat{M}^{A}_{I_{l}}\hat{M}^{B}_{J_{l'}}\partial^{A}_{I_{l}J_{l'}}\left(\frac{1}{|\boldsymbol{x}^{A}-\boldsymbol{x}^{B}|}\right)\notag\\
&&-\frac{G}{6}\sum_{A=1}^{N}\sum_{B\neq A}\sum_{l=0}^{\infty}\sum_{l'=0}^{\infty}\frac{(-1)^{l'}}{l!\,l'!}\hat{Q}^{A}_{I_{l}}\hat{Q}^{B}_{J_{l'}}\partial^{A}_{I_{l}J_{l'}}\left(\frac{\text{e}^{-m_{\text{s}}|\boldsymbol{x}^{A}-\boldsymbol{x}^{B}|}}{|\boldsymbol{x}^{A}-\boldsymbol{x}^{B}|}\right).
\end{eqnarray}
This expression demonstrates that the total energy of the system comprises the self-energies, the center-of-mass kinetic energies, and the mutual interaction potential energies of all the bodies, which is similar to the case in GR. However, despite this similarity, both the self-energies and the mutual interaction potential energies of the bodies in Eq.~(\ref{equ4.39}) differ significantly from their counterparts in GR due to the distinct gravitational potential produced by a gravitating source in $f(R)$ gravity compared to GR. For the system, the assumption that the typical
size of each body is far less than the typical separation between bodies implies that each body can be approximately treated as a self-gravitating system, so the self-energy of each body remains approximately conserved, which can also be immediately verified by following the analogous process outlined in Eqs.~(\ref{equ4.18})--(\ref{equ4.29}).  In light of this, we shall no longer focus on the self-energies of bodies. In addition, since the center-of-mass kinetic energies of bodies are conventionally defined, our attention will turn to the interaction potential energies between bodies in what follows.
Equation~(\ref{equ4.39}) indicates that the interaction potential energies between different bodies within the framework of $f(R)$ gravity comprise two components: one arising from couplings of mass multipole moments, and the other resulting from the couplings of scalar multipole moments.
To better illustrate the roles of the potential energies due to multipole moment couplings between bodies, as in Eq.~(\ref{equ3.65}), the above expression for the total energy of the system is rearranged as follows:
\begin{eqnarray}
E&=&\sum_{A=1}^{N}\frac{1}{2}m^{A}\left(\boldsymbol{v}^{A}\right)^{2}-\frac{1}{2}\sum_{A=1}^{N}\sum_{B\neq A}\frac{Gm^{A}m^{B}}{|\boldsymbol{x}^{A}-\boldsymbol{x}^{B}|}-\frac{1}{6}\sum_{A=1}^{N}\sum_{B\neq A}\frac{GQ^{A}Q^{B}}{|\boldsymbol{x}^{A}-\boldsymbol{x}^{B}|}\text{e}^{-m_{\text{s}}|\boldsymbol{x}^{A}-\boldsymbol{x}^{B}|}\notag\\
&&-\frac{G}{2}\sum_{A=1}^{N}\sum_{B\neq A}\sum_{l=2}^{\infty}\frac{1}{l!}\left[(-1)^{l}m^{A}\hat{M}^{B}_{I_{l}}+\hat{M}^{A}_{I_{l}}m^{B}\right]\partial^{A}_{I_{l}}\left(\frac{1}{|\boldsymbol{x}^{A}-\boldsymbol{x}^{B}|}\right)\notag
\end{eqnarray}
\begin{eqnarray}
&&-\frac{G}{2}\sum_{A=1}^{N}\sum_{B\neq A}\sum_{l=2}^{\infty}\sum_{l'=2}^{\infty}\frac{(-1)^{l'}}{l!\,l'!}\hat{M}^{A}_{I_{l}}\hat{M}^{B}_{J_{l'}}\partial^{A}_{I_{l}J_{l'}}\left(\frac{1}{|\boldsymbol{x}^{A}-\boldsymbol{x}^{B}|}\right)\notag\\
&&-\frac{G}{6}\sum_{A=1}^{N}\sum_{B\neq A}\sum_{l=1}^{\infty}\frac{1}{l!}\left[(-1)^{l}Q^{A}\hat{Q}^{B}_{I_{l}}+\hat{Q}^{A}_{I_{l}}Q^{B}\right]\partial^{A}_{I_{l}}\left(\frac{\text{e}^{-m_{\text{s}}|\boldsymbol{x}^{A}-\boldsymbol{x}^{B}|}}{|\boldsymbol{x}^{A}-\boldsymbol{x}^{B}|}\right)\notag\\
\label{equ4.40}
&&-\frac{G}{6}\sum_{A=1}^{N}\sum_{B\neq A}\sum_{l=1}^{\infty}\sum_{l'=1}^{\infty}\frac{(-1)^{l'}}{l!\,l'!}\hat{Q}^{A}_{I_{l}}\hat{Q}^{B}_{J_{l'}}\partial^{A}_{I_{l}J_{l'}}\left(\frac{\text{e}^{-m_{\text{s}}|\boldsymbol{x}^{A}-\boldsymbol{x}^{B}|}}{|\boldsymbol{x}^{A}-\boldsymbol{x}^{B}|}\right),
\end{eqnarray}
where the self-energies of bodies have been omitted. As mentioned previously, all the non-monopole moments of an ideal spherical body vanish, which implies that if all bodies in the system are ideal spheres, the total energy of the system simplifies to the first line in Eq.~(\ref{equ4.40}) (the monopole-monopole potential energy), namely,
\begin{eqnarray}
\label{equ4.41}E&=&\sum_{A=1}^{N}\frac{1}{2}m^{A}\left(\boldsymbol{v}^{A}\right)^{2}-\frac{1}{2}\sum_{A=1}^{N}\sum_{B\neq A}\frac{Gm^{A}m^{B}}{|\boldsymbol{x}^{A}-\boldsymbol{x}^{B}|}-\frac{1}{6}\sum_{A=1}^{N}\sum_{B\neq A}\frac{GQ^{A}Q^{B}}{|\boldsymbol{x}^{A}-\boldsymbol{x}^{B}|}\text{e}^{-m_{\text{s}}|\boldsymbol{x}^{A}-\boldsymbol{x}^{B}|},\qquad\quad
\end{eqnarray}
where the last term represents the modification to the familiar expression in classical celestial mechanics under $f(R)$ gravity. A comparison between this result and the reduced version for a point-body system~(cf.~Eq.~(\ref{equ3.63}))
\begin{eqnarray}
\label{equ4.42}E&=&\sum_{A=1}^{N}\frac{1}{2}m^{A}\left(\boldsymbol{v}^{A}\right)^{2}-\frac{1}{2}\sum_{A=1}^{N}\sum_{B\neq A}\frac{Gm^{A}m^{B}}{|\boldsymbol{x}^{A}-\boldsymbol{x}^{B}|}-\frac{1}{6}\sum_{A=1}^{N}\sum_{B\neq A}\frac{Gm^{A}m^{B}}{|\boldsymbol{x}^{A}-\boldsymbol{x}^{B}|}\text{e}^{-m_{\text{s}}|\boldsymbol{x}^{A}-\boldsymbol{x}^{B}|}\qquad\quad
\end{eqnarray}
further highlights that for a system composed of multiple ideal spheres, the sizes of the bodies play a crucial role in determining the total energy of the system, which is different from the cases in GR and massless scalar-tensor theories. In GR, the total energy of a system composed of ideal spheres is independent of the sizes of the bodies. However, such a conclusion does not universally hold in massless scalar-tensor theories, because the structure of multipole moments in these theories is more subtle, and the notion of sphericity itself depends on the specific definition of the multipole moments.  Let us return to the system of non-spherical bodies. The remaining lines in Eq.~(\ref{equ4.40}) suggest that apart from the monopole-monopole potential energy, the total energy of such system is also contributed by all the monopole-multipole and multipole-multipole potential energies between different bodies. Since the total energy of the system in GR can be directly obtained when all the scalar multipole moments of bodies vanish, from the remaining lines in Eq.~(\ref{equ4.40}), it is clear that the monopole-dipole and dipole-dipole potential energies do not contribute to the interaction potential energies of bodies in GR. However, these potential energies are significant in $f(R)$ gravity, as the scalar dipole moments of bodies remain non-zero. Similar dipole couplings also appear in massless scalar-tensor theories, but there the scalar field is long-ranged. In $f(R)$ gravity, by contrast, the Yukawa suppression leads to the massive scalar field being short-ranged. These observations highlight key distinctions between $f(R)$ gravity and both GR and massless scalar-tensor theories. Finally, from Eq.~(\ref{equ4.40}), we observe that the interaction potential energies of bodies depend on the multi-index derivatives of $1/|\boldsymbol{x}^{A}-\boldsymbol{x}^{B}|$ and $\text{e}^{-m_{\text{s}}|\boldsymbol{x}^{A}-\boldsymbol{x}^{B}|}/|\boldsymbol{x}^{A}-\boldsymbol{x}^{B}|$, and it follows that as the typical separation $|\boldsymbol{x}^{A}-\boldsymbol{x}^{B}|$ between bodies approaches positive infinity, these potential energies vanish, which precisely captures the essential nature of the interaction potential energy.
\subsection{Spin dynamics in the system}
As previously noted, in an isolated self-gravitating system, the conservation of total angular momentum ensures that the combined orbital and spin angular momenta of all bodies remain invariant. This observation suggests that the spin angular momentum of an
individual body is not generally conserved due to the interactions with other bodies within the system. Specifically, in this subsection, we intend to analyze the temporal evolution of each body's spin angular momentum and establish the associated
equation of motion in $f(R)$ gravity at the Newtonian order.

For body $A$, as in Eq.~(\ref{equ4.9}), its spin angular momentum is given by
\begin{eqnarray}
\label{equ4.43}&&\boldsymbol{S}^{A}(t):=\int_{A}\rho(t,\boldsymbol{x})\left(\boldsymbol{x}-\boldsymbol{x}^{A}\right)\times\left(\boldsymbol{v}-\boldsymbol{v}^{A}\right)\,\text{d}^{3}x,
\end{eqnarray}
and by exploiting the formula~(\ref{equ3.12}) once again, the differentiation of $\boldsymbol{S}^{A}$ with respect to time is evaluated as
\begin{eqnarray}
\label{equ4.44}\frac{\text{d}\boldsymbol{S}^{A}}{\text{d}t}&=&\int_{A}\rho\left(\boldsymbol{x}-\boldsymbol{x}^{A}\right)\times\left(\frac{\text{d}\boldsymbol{v}}{\text{d}t}-\boldsymbol{a}^{A}\right)\,\text{d}^{3}x=\int_{A}\rho\left(\boldsymbol{x}-\boldsymbol{x}^{A}\right)\times\frac{\text{d}\boldsymbol{v}}{\text{d}t}\,\text{d}^{3}x\notag\\
&=&\int_{A}\rho\left(\boldsymbol{x}-\boldsymbol{x}^{A}\right)\times\nabla\varPhi\,\text{d}^{3}x-\int_{A}\left(\boldsymbol{x}-\boldsymbol{x}^{A}\right)\times\nabla p\,\text{d}^{3}x\notag\\
&=&\int_{A}\rho\left(\boldsymbol{x}-\boldsymbol{x}^{A}\right)\times\nabla\varPhi\,\text{d}^{3}x,
\end{eqnarray}
where the definitions of the center-of-mass position and velocity are used in the first and second steps, the Euler's equation~(\ref{equ3.5}) is inserted in the third step, and the following identity
\begin{eqnarray}
\label{equ4.45}&&\int_{A}\nabla\times\left[p\left(\boldsymbol{x}-\boldsymbol{x}^{A}\right)\right]\,\text{d}^{3}x=\oint_{\partial A}\text{d}\boldsymbol{S}\times\left[p\left(\boldsymbol{x}-\boldsymbol{x}^{A}\right)\right]=0
\end{eqnarray}
is invoked in the last step due to the vanishing pressure on the boundary $\partial{A}$ of the integration region. By substituting the decomposition of the gravitational potential given in Eq.~(\ref{equ3.15}), the above expression for $\text{d}\boldsymbol{S}^{A}/\text{d}t$ becomes
\begin{eqnarray}
\label{equ4.46}\frac{\text{d}\boldsymbol{S}^{A}}{\text{d}t}
&=&\int_{A}\rho\left(\boldsymbol{x}-\boldsymbol{x}^{A}\right)\times\nabla\varPhi^{A}\,\text{d}^{3}x+\int_{A}\rho\left(\boldsymbol{x}-\boldsymbol{x}^{A}\right)\times\nabla\varPhi^{-A}\,\text{d}^{3}x.
\end{eqnarray}
Because of Eqs.~(\ref{equ3.17}) and~(\ref{equ3.2})--(\ref{equ3.4}) , the first term is simplified to be
\begin{eqnarray}
\label{equ4.47}&&\int_{A}\rho\left(\boldsymbol{x}-\boldsymbol{x}^{A}\right)\times\nabla\varPhi^{A}\,\text{d}^{3}x=\int_{A}\rho\boldsymbol{x}\times\nabla\varPhi^{A}\,\text{d}^{3}x\notag\\
&=&G\int_{A}\int_{A}\rho\rho'\boldsymbol{x}\times\nabla\left(\frac{1}{|\boldsymbol{x}-\boldsymbol{x}'|}\right)\,\text{d}^{3}x\text{d}^{3}x'+\frac{G}{3}\int_{A}\int_{A}\rho\rho'\boldsymbol{x}\times\nabla\left(\frac{\text{e}^{-m_{\text{s}}|\boldsymbol{x}-\boldsymbol{x}'|}}{|\boldsymbol{x}-\boldsymbol{x}'|}\right)\,\text{d}^{3}x\text{d}^{3}x'\notag\\
&=&0.
\end{eqnarray}
The result in the last step is obtained by applying the derivation technique introduced in Eqs.~(\ref{equ4.12}) and (\ref{equ4.13}). Thus, the expression for $\text{d}\boldsymbol{S}^{A}/\text{d}t$ should be
\begin{eqnarray}
\label{equ4.48}\frac{\text{d}\boldsymbol{S}^{A}}{\text{d}t}
&=&\int_{A}\rho\left(\boldsymbol{x}-\boldsymbol{x}^{A}\right)\times\nabla\varPhi^{-A}\,\text{d}^{3}x,
\end{eqnarray}
which explicitly suggests that analogous to the case of the center-of-center acceleration, the time derivative of $\boldsymbol{S}^{A}$ is determined solely by the external potential generated by other bodies in the system. In order to conveniently perform the multipole expansion for $\text{d}\boldsymbol{S}^{A}/\text{d}t$ later, we write
\begin{eqnarray}
\label{equ4.49}\frac{\text{d}S_{i}^{A}}{\text{d}t}
&=&\epsilon_{ijk}\int_{A}\rho\left(x-x^{A}\right)_{j}\partial_{k}\varPhi^{-A}\,\text{d}^{3}x.
\end{eqnarray}
Given that the external potential $\varPhi^{-A}$ can also be decomposed into the external Coulomb-like potential $U^{-A}$ and the external Yukawa-like potential $Y^{-A}$,
from their expressions in Eqs.~(\ref{equ3.22}) and (\ref{equ3.23}), we have
\begin{eqnarray}
\label{equ4.50}\frac{\text{d}S_{i}^{A}}{\text{d}t}
&=&G\epsilon_{ijk}\sum_{B\neq A}\int_{A}\int_{B}\rho\rho'\left(x-x^{A}\right)_{j}\frac{\partial}{\partial x_{k}}\left(\frac{1}{|\boldsymbol{x}-\boldsymbol{x}'|}\right)\,\text{d}^{3}x\text{d}^{3}x'\notag\\
&&+\frac{G}{3}\epsilon_{ijk}\sum_{B\neq A}\int_{A}\int_{B}\rho\rho'\left(x-x^{A}\right)_{j}\frac{\partial}{\partial x_{k}}\left(\frac{\text{e}^{-m_{\text{s}}|\boldsymbol{x}-\boldsymbol{x}'|}}{|\boldsymbol{x}-\boldsymbol{x}'|}\right)\,\text{d}^{3}x\text{d}^{3}x'.\qquad
\end{eqnarray}
The multipole expansions for the spatial derivatives of the functions $1/|\boldsymbol{x}-\boldsymbol{x}'|$ and $\text{e}^{-m_{\text{s}}|\boldsymbol{x}-\boldsymbol{x}'|}/$ $|\boldsymbol{x}-\boldsymbol{x}'|$ have been provided in Eqs.~(\ref{equ3.51}) and (\ref{equ3.52}), so
replacing them into the above two equations yields the multipole expansion for $\text{d}S_{i}^{A}/\text{d}t$,
\begin{eqnarray}
\label{equ4.51}\frac{\text{d}S_{i}^{A}}{\text{d}t}
&=&G\epsilon_{ijk}\sum_{B\neq A}\sum_{l=0}^{\infty}\sum_{l'=0}^{\infty}\frac{(-1)^{l'}}{l!\,l'!}\hat{M}^{A}_{jI_{l}}\hat{M}^{B}_{J_{l'}}\partial^{A}_{kI_{l}J_{l'}}\left(\frac{1}{|\boldsymbol{x}^{A}-\boldsymbol{x}^{B}|}\right)\notag\\
&&+\frac{G}{3}\epsilon_{ijk}\sum_{B\neq A}\sum_{l=0}^{\infty}\sum_{l'=0}^{\infty}\frac{(-1)^{l'}}{l!\,l'!}\int_{A}\rho\left(x-x^{A}\right)_{\langle jI_{l}\rangle}\updelta_{l}\left(m_{\text{s}}|\boldsymbol{x}-\boldsymbol{x}^{A}|\right)\text{d}^{3}x\,\hat{Q}^{B}_{J_{l'}}\, \partial^{A}_{kI_{l}J_{l'}}\left(\frac{\text{e}^{-m_{\text{s}}|\boldsymbol{x}^{A}-\boldsymbol{x}^{B}|}}{|\boldsymbol{x}^{A}-\boldsymbol{x}^{B}|}\right).
\end{eqnarray}
In the above derivations, two critical issues need to be addressed. One is that there exists
\begin{eqnarray}
\label{equ4.52}\left(x-x^{A}\right)_{j}\left(x-x^{A}\right)_{\langle I_{l}\rangle}=\left(x-x^{A}\right)_{j\langle I_{l}\rangle}=\left(x-x^{A}\right)_{\langle jI_{l}\rangle},
\end{eqnarray}
which holds because when $i_{k}=j\ (k=1,2,\cdots l)$, the derivative operator $\partial^{A}_{kI_{l}J_{l'}}$ reduces to $\partial^{A}_{kjI_{l-1}J_{l'}}$, and the result vanishes due to the multiplication by the totally antisymmetric Levi-Civita tensor $\epsilon_{ijk}$. Another is the appearance of the integral $\int_{A}\rho\left(x-x^{A}\right)_{\langle jI_{l}\rangle}\updelta_{l}\left(m_{\text{s}}|\boldsymbol{x}-\boldsymbol{x}^{A}|\right)\text{d}^{3}x$, which differs from the $(l+1)$-order scalar multipole moment of body $A$, as defined in Eq.~(\ref{equ3.57}). Despite this, in view that the integral still describes the mass distribution of body $A$, it should also be identified as some type of multipole moment. In addition, the structural similarity between this integral and that one in the definition of scalar multipole moment strongly implies that the concept of the scalar multipole moment needs to be extended. To this end, we define the scalar multipole moments with weight $k$ as
\begin{eqnarray}
\label{equ4.53}&&\hat{Q}^{A[k]}_{I_{l}}(t):=\int_{A}\rho(t,\boldsymbol{x})\left(x-x^{A}\right)_{\langle I_{l}\rangle}\updelta_{k}\left(m_{\text{s}}|\boldsymbol{x}-\boldsymbol{x}^{A}|\right)\text{d}^{3}x,
\end{eqnarray}
from which, one can infer that the previous $l$-order scalar multipole moment corresponds to the one with weight $l$. The introduction of the weighted scalar multipole moment allows us to recast the multipole expansion of $\text{d}S_{i}^{A}/\text{d}t$ as
\begin{eqnarray}
\label{equ4.54}\frac{\text{d}S_{i}^{A}}{\text{d}t}
&=&G\epsilon_{ijk}\sum_{B\neq A}\sum_{l=0}^{\infty}\sum_{l'=0}^{\infty}\frac{(-1)^{l'}}{l!\,l'!}\hat{M}^{A}_{jI_{l}}\hat{M}^{B}_{J_{l'}}\partial^{A}_{kI_{l}J_{l'}}\left(\frac{1}{|\boldsymbol{x}^{A}-\boldsymbol{x}^{B}|}\right)\notag\\
&&+\frac{G}{3}\epsilon_{ijk}\sum_{B\neq A}\sum_{l=0}^{\infty}\sum_{l'=0}^{\infty}\frac{(-1)^{l'}}{l!\,l'!}\hat{Q}^{A[l]}_{jI_{l}}\hat{Q}^{B[l']}_{J_{l'}}\partial^{A}_{kI_{l}J_{l'}}\left(\frac{\text{e}^{-m_{\text{s}}|\boldsymbol{x}^{A}-\boldsymbol{x}^{B}|}}{|\boldsymbol{x}^{A}-\boldsymbol{x}^{B}|}\right).\qquad
\end{eqnarray}
As in the case for the center-of-center acceleration $a^{A}_{i}$, this equation shows that the multipole expansion of $\text{d}S_{i}^{A}/\text{d}t$ also comprises the Coulomb-type part (corresponding to the result in GR~\cite{Eric2014}) and the Yukawa-type part (representing the modification introduced by $f(R)$ gravity). However, due to the presence of the index $j$, the products of multipole moments appearing in the two parts differ from their counterparts in the multipole expansion of $a^{A}_{i}$. Specifically, one can observe that the product of monopole moments makes no contribution to the motion of $\boldsymbol{S}^{A}$, which reflects an obvious distinction between spin and translational acceleration. To better understand the roles of the multipole moment products, Eq.~(\ref{equ4.54}) can be rewritten in a more accessible form as follows:
\begin{eqnarray}
\label{equ4.55}\frac{\text{d}S_{i}^{A}}{\text{d}t}
&=&G\epsilon_{ijk}\sum_{B\neq A}\sum_{l=1}^{\infty}\frac{1}{l!}\hat{M}^{A}_{jI_{l}}m^{B}\partial^{A}_{kI_{l}}\left(\frac{1}{|\boldsymbol{x}^{A}-\boldsymbol{x}^{B}|}\right)\notag\\
&&+G\epsilon_{ijk}\sum_{B\neq A}\sum_{l=1}^{\infty}\sum_{l'=2}^{\infty}\frac{(-1)^{l'}}{l!\,l'!}\hat{M}^{A}_{jI_{l}}\hat{M}^{B}_{J_{l'}}\partial^{A}_{kI_{l}J_{l'}}\left(\frac{1}{|\boldsymbol{x}^{A}-\boldsymbol{x}^{B}|}\right)\notag\\
&&+\frac{G}{3}\epsilon_{ijk}\sum_{B\neq A}\sum_{l=0}^{\infty}\frac{1}{l!}\hat{Q}^{A[l]}_{jI_{l}}Q^{B}\partial^{A}_{kI_{l}}\left(\frac{\text{e}^{-m_{\text{s}}|\boldsymbol{x}^{A}-\boldsymbol{x}^{B}|}}{|\boldsymbol{x}^{A}-\boldsymbol{x}^{B}|}\right)\notag\\
&&+\frac{G}{3}\epsilon_{ijk}\sum_{B\neq A}\sum_{l=0}^{\infty}\sum_{l'=1}^{\infty}\frac{(-1)^{l'}}{l!\,l'!}\hat{Q}^{A[l]}_{jI_{l}}\hat{Q}^{B[l']}_{J_{l'}}\partial^{A}_{kI_{l}J_{l'}}\left(\frac{\text{e}^{-m_{\text{s}}|\boldsymbol{x}^{A}-\boldsymbol{x}^{B}|}}{|\boldsymbol{x}^{A}-\boldsymbol{x}^{B}|}\right),\qquad
\end{eqnarray}
where those terms involving the monopole moment of body $B$ have been isolated, and the conclusion that the mass dipole moments of bodies vanish is employed. It is shown that for body $A$, the lack of $m^{A}$ and $Q^{A}$ implies that its own monopole moments do not affect the motion of $\boldsymbol{S}^{A}$, whereas those terms involving $\hat{M}^{A}_{jI_{l}}m^{B}$ and $\hat{Q}^{A[l]}_{jI_{l}}Q^{B}$ reveal that its higher-order multipole moments can affect the motion of $\boldsymbol{S}^{A}$, which is consistent with the inherent properties of the spin angular momentum. Combined with the result from equation (\ref{equ3.65}), this observation highlights that both the size and shape of a body not only impact its orbital motion but also significantly influence its spin motion in $f(R)$ gravity. The remaining terms in Eq.~(\ref{equ4.55}) suggest that the couplings between the higher-order multipole moments of body $A$ and those of other bodies can also influence the motion of $\boldsymbol{S}^{A}$. Although this is analogous to the case for the center-of-center acceleration $\boldsymbol{a}^{A}$, it should be noted that the scalar multipole moments involved in the multipole expansions of $S_{i}^{A}$ and $a^{A}_{i}$ differ, where the definition of those in the expansion of $S_{i}^{A}$ has been slightly extended as outlined in Eq.~(\ref{equ4.53}). Finally, from the terms in the Coulomb-type and Yukawa-type parts, we recognize that the mass dipole moment of body $A$ fails to contribute to the motion of $\boldsymbol{S}^{A}$, whereas its scalar dipole moment plays a role. While similar scalar dipole effects are known in the context of massless scalar-tensor theories (e.g., Ref.~\cite{Kopeikin2019}), in $f(R)$ gravity these interactions are uniquely governed by the massive Yukawa sector, a feature that distinguishes it from both GR and the massless case.

Equation~(\ref{equ4.55}), being the equation of motion for $\boldsymbol{S}^{A}$, describes how the spin angular momentum of each body evolves over time and thereby determines the behavior of each spin in $f(R)$ gravity at the Newtonian order. Explicitly, once the mass and scalar multipole moments of the bodies are specified based on their internal dynamics, the multipole expansions for the time derivatives of the spin angular momenta of all the bodies form a complete set of equations for the spin dynamics of the system. Consequently, just like Eq.~(\ref{equ3.65}), this equation could serve as the starting point for understanding the spin motion of the system within $f(R)$ gravity.
\section{Orbital motion of the two-body system~\label{Sec:fifth}}
In the Universe, the most common many-body system is the two-body system, and such a typical example is the binary system of main-sequence stars. The application of this paper's previous findings to studying the orbital dynamics of a two-body system is  pivotal for testing $f(R)$ gravity. Just as demonstrated in classical celestial mechanics, the orbital dynamics of a two-body system are often investigated using the effective one-body approach. To facilitate future applications, this section will be dedicated to deriving the effective one-body equation governing a two-body system and establishing the expression for the total energy of this system in the Newtonian limit of $f(R)$ gravity.

Consider a two-body system consisting of a first body of mass $m^{1}$ at a position $\boldsymbol{x}^{1}$ and a second body of mass $m^{2}$ at a position $\boldsymbol{x}^{2}$. According to Eq.~(\ref{equ4.6}), the barycenter position of the two-body system is located at
\begin{eqnarray}
\label{equ5.1}&&\boldsymbol{X}:=\frac{1}{m}\left(m^{1}\boldsymbol{x}^{1}+m^{2}\boldsymbol{x}^{2}\right)
\end{eqnarray}
with $m:=m^{1}+m^{2}$ as the total mass of the system. In the spirit of the effective one-body approach, the origin of the coordinate system is chosen to coincide with the barycenter, and the relative motion between the two bodies needs to be analyzed. With these requirements, the relative separation between the bodies can be defined as
\begin{eqnarray}
\label{equ5.2}&&\boldsymbol{x}=\boldsymbol{x}^{1}-\boldsymbol{x}^{2},
\end{eqnarray}
and by using the identity
\begin{eqnarray}
\label{equ5.3}&&m^{1}\boldsymbol{x}^{1}+m^{2}\boldsymbol{x}^{2}=0,
\end{eqnarray}
we obtain
\begin{eqnarray}
\label{equ5.4}&&\boldsymbol{x}^{1}=\frac{m^{2}}{m}\boldsymbol{x},\qquad \boldsymbol{x}^{2}=-\frac{m^{1}}{m}\boldsymbol{x}.
\end{eqnarray}
These results indicate that once the relative separation $\boldsymbol{x}$ of the two bodies is given, their orbital motions are fully determined, and hence, our subsequent focus will turn to studying the relative motion between them.
In the framework of Newtonian $f(R)$ gravity (i.e.,~$f(R)$ gravity at the Newtonian order), the relative velocity and acceleration between them should be evaluated by
\begin{eqnarray}
\label{equ5.5}&&\boldsymbol{v}:=\frac{\text{d}\boldsymbol{x}}{\text{d}t}=\boldsymbol{v}^{1}-\boldsymbol{v}^{2},\\
\label{equ5.6}&&\boldsymbol{a}:=\frac{\text{d}\boldsymbol{v}}{\text{d}t}=\boldsymbol{a}^{1}-\boldsymbol{a}^{2},
\end{eqnarray}
where $\boldsymbol{v}^{1}=\text{d}\boldsymbol{x}^{1}/\text{d}t$ and $\boldsymbol{v}^{2}=\text{d}\boldsymbol{x}^{2}/\text{d}t$ are their velocities, and $\boldsymbol{a}^{1}=\text{d}\boldsymbol{v}^{1}/\text{d}t$ and $\boldsymbol{a}^{2}=\text{d}\boldsymbol{v}^{2}/\text{d}t$ are their accelerations. Clearly, the relative motion is determined by $\boldsymbol{a}$, and since the multipole expansions of $\boldsymbol{a}^{1}$ and $\boldsymbol{a}^{2}$ have been provided
in Eq.~(\ref{equ3.65}), the corresponding expansion for the components of $\boldsymbol{a}$ can be derived as
\begin{eqnarray}
a_{i}&=&-\frac{Gm}{|\boldsymbol{x}|^{3}}x_{i}+\frac{G}{3\mu}Q^{1}Q^{2}\partial_{i}\left(\frac{\text{e}^{-m_{\text{s}}|\boldsymbol{x}|}}{|\boldsymbol{x}|}\right)\notag\\
&&+\frac{G}{\mu}\sum_{l=2}^{\infty}\frac{1}{l!}\left[m^{2}\hat{M}^{1}_{I_{l}}+(-1)^{l}m^{1}\hat{M}^{2}_{I_{l}}\right]\partial_{iI_{l}}\left(\frac{1}{|\boldsymbol{x}|}\right)\notag
\end{eqnarray}
\begin{eqnarray}
&&+\frac{G}{\mu}\sum_{l=2}^{\infty}\sum_{l'=2}^{\infty}\frac{(-1)^{l'}}{l!\,l'!}\hat{M}^{1}_{I_{l}}\hat{M}^{2}_{J_{l'}}\partial_{iI_{l}J_{l'}}\left(\frac{1}{|\boldsymbol{x}|}\right)\notag\\
&&+\frac{G}{3\mu}\sum_{l=1}^{\infty}\frac{1}{l!}\left[Q^{2}\hat{Q}^{1}_{I_{l}}+(-1)^{l}Q^{1}\hat{Q}^{2}_{I_{l}}\right]\partial_{iI_{l}}\left(\frac{\text{e}^{-m_{\text{s}}|\boldsymbol{x}|}}{|\boldsymbol{x}|}\right)\notag\\
\label{equ5.7}&&+\frac{G}{3\mu}\sum_{l=1}^{\infty}\sum_{l'=1}^{\infty}\frac{(-1)^{l'}}{l!\,l'!}\hat{Q}^{1}_{I_{l}}\hat{Q}^{2}_{J_{l'}}\partial_{iI_{l}J_{l'}}\left(\frac{\text{e}^{-m_{\text{s}}|\boldsymbol{x}|}}{|\boldsymbol{x}|}\right).
\end{eqnarray}
This is exactly the effective one-body equation governing the relative motion between the two bodies in $f(R)$ gravity at the Newtonian order. It is important to note that in this equation, $\mu:=m^{1}m^{2}/m$ is the reduced mass of the two-body system, $x_{i}$ denotes a component of the relative separation $\boldsymbol{x}$, and $\partial_{i}:=\partial/\partial x_{i}$. The above derivation relies on the following two basic facts that
\begin{eqnarray}
\label{equ5.8}&&\partial^{1}_{i}\left(\frac{1}{|\boldsymbol{x}|}\right)=-\partial^{2}_{i}\left(\frac{1}{|\boldsymbol{x}|}\right)=\partial_{i}\left(\frac{1}{|\boldsymbol{x}|}\right),\\
\label{equ5.9}&&\partial^{1}_{i}\left(\frac{\text{e}^{-m_{\text{s}}|\boldsymbol{x}|}}{|\boldsymbol{x}|}\right)=-\partial^{2}_{i}\left(\frac{\text{e}^{-m_{\text{s}}|\boldsymbol{x}|}}{|\boldsymbol{x}|}\right)=\partial_{i}\left(\frac{\text{e}^{-m_{\text{s}}|\boldsymbol{x}|}}{|\boldsymbol{x}|}\right).
\end{eqnarray}
From Eqs.~(\ref{equ5.4}), we find that the velocities of the two bodies satisfy the following identities
 \begin{eqnarray}
\label{equ5.10}&&\boldsymbol{v}^{1}=\frac{m^{2}}{m}\boldsymbol{v},\qquad \boldsymbol{v}^{2}=-\frac{m^{1}}{m}\boldsymbol{v},
\end{eqnarray}
and based on them, by applying the facts~(\ref{equ5.8}) and (\ref{equ5.9}) again, the total energy of the two-body system in $f(R)$ gravity at the Newtonian order can also be derived from Eq.~(\ref{equ4.40}), namely,
\begin{eqnarray}
\label{equ5.11}
E&=&\frac{1}{2}\mu\boldsymbol{v}^{2}-\frac{Gm\mu}{|\boldsymbol{x}|}-\frac{GQ^{1}Q^{2}}{3}\frac{\text{e}^{-m_{\text{s}}|\boldsymbol{x}|}}{|\boldsymbol{x}|}\notag\\
&&-G\sum_{l=2}^{\infty}\frac{1}{l!}\left[(-1)^{l}m^{1}\hat{M}^{2}_{I_{l}}+\hat{M}^{1}_{I_{l}}m^{2}\right]\partial_{I_{l}}\left(\frac{1}{|\boldsymbol{x}|}\right)\notag\\
&&-G\sum_{l=2}^{\infty}\sum_{l'=2}^{\infty}\frac{(-1)^{l'}}{l!\,l'!}\hat{M}^{1}_{I_{l}}\hat{M}^{2}_{J_{l'}}\partial_{I_{l}J_{l'}}\left(\frac{1}{|\boldsymbol{x}|}\right)\notag\\
&&-\frac{G}{3}\sum_{l=1}^{\infty}\frac{1}{l!}\left[(-1)^{l}Q^{1}\hat{Q}^{2}_{I_{l}}+\hat{Q}^{1}_{I_{l}}Q^{2}\right]\partial_{I_{l}}\left(\frac{\text{e}^{-m_{\text{s}}|\boldsymbol{x}|}}{|\boldsymbol{x}|}\right)\notag\\
&&-\frac{G}{3}\sum_{l=1}^{\infty}\sum_{l'=1}^{\infty}\frac{(-1)^{l'}}{l!\,l'!}\hat{Q}^{1}_{I_{l}}\hat{Q}^{2}_{J_{l'}}\partial_{I_{l}J_{l'}}\left(\frac{\text{e}^{-m_{\text{s}}|\boldsymbol{x}|}}{|\boldsymbol{x}|}\right).
\end{eqnarray}

Equations~(\ref{equ5.7}) and (\ref{equ5.11}) constitute the basis for investigating the orbital motion of a two-body system in $f(R)$ gravity. If both the bodies are ideal spheres, all their higher-order multipole moments vanish, which leads to the degeneration of Eqs.~(\ref{equ5.7}) and (\ref{equ5.11}) into
\begin{eqnarray}
\label{equ5.12}&&a_{i}=-\frac{Gm}{|\boldsymbol{x}|^{3}}x_{i}+\frac{G}{3\mu}Q^{1}Q^{2}\partial_{i}\left(\frac{\text{e}^{-m_{\text{s}}|\boldsymbol{x}|}}{|\boldsymbol{x}|}\right),\\
\label{equ5.13}&&E=\frac{1}{2}\mu\boldsymbol{v}^{2}-\frac{Gm\mu}{|\boldsymbol{x}|}-\frac{GQ^{1}Q^{2}}{3}\frac{\text{e}^{-m_{\text{s}}|\boldsymbol{x}|}}{|\boldsymbol{x}|}.
\end{eqnarray}
It is not difficult to identify that the above two equations explicitly yield the modifications to the corresponding results in GR. Starting from them, one can systematically explore the unperturbed orbits of two bodies within Newtonian $f(R)$ gravity for systems composed of spherical bodies. As previously noted, the presence of the scalar monopole moments in these two equations suggests that the mass distributions of the bodies play significant roles in shaping their orbits. Furthermore, when the two bodies are sufficiently far apart that they can be treated as point particles, due to Eq.~(\ref{equ3.63}), Eqs.~(\ref{equ5.12}) and (\ref{equ5.13}) reduce to
\begin{eqnarray}
\label{equ5.14}&&a_{i}=-\frac{Gm}{|\boldsymbol{x}|^{3}}x_{i}+\frac{Gm}{3}\partial_{i}\left(\frac{\text{e}^{-m_{\text{s}}|\boldsymbol{x}|}}{|\boldsymbol{x}|}\right),\\
\label{equ5.15}&&E=\frac{1}{2}\mu\boldsymbol{v}^{2}-\frac{Gm\mu}{|\boldsymbol{x}|}-\frac{Gm\mu}{3}\frac{\text{e}^{-m_{\text{s}}|\boldsymbol{x}|}}{|\boldsymbol{x}|}.
\end{eqnarray}
With these equations, we can fully determine the unperturbed orbits of a system comprising two point particles, even when the masses of the bodies are comparable, and these orbits represent generalizations of their counterparts (viz. conic sections) in Newtonian gravity.
In addition to the application to spherical bodies, Eqs.~(\ref{equ5.7}) and (\ref{equ5.11}) can also be effectively utilized in cases where the higher-order multipole moments of body 1 are negligible.  Such a typical example is a planet orbiting the Sun, where all the higher-order multipole moments of the planet are approximately zero. Under this circumstance, Eqs.~(\ref{equ5.7}) and (\ref{equ5.11}) are, respectively, simplified to be
\begin{eqnarray}
\label{equ5.16}a_{i}&=&-\frac{Gm}{|\boldsymbol{x}|^{3}}x_{i}+\frac{G}{3\mu}Q^{1}Q^{2}\partial_{i}\left(\frac{\text{e}^{-m_{\text{s}}|\boldsymbol{x}|}}{|\boldsymbol{x}|}\right)+\frac{G}{\mu}\sum_{l=2}^{\infty}\frac{(-1)^{l}}{l!}m^{1}\hat{M}^{2}_{I_{l}}\partial_{iI_{l}}\left(\frac{1}{|\boldsymbol{x}|}\right)\notag\\
&&+\frac{G}{3\mu}\sum_{l=1}^{\infty}\frac{(-1)^{l}}{l!}Q^{1}\hat{Q}^{2}_{I_{l}}\partial_{iI_{l}}\left(\frac{\text{e}^{-m_{\text{s}}|\boldsymbol{x}|}}{|\boldsymbol{x}|}\right),\\
\label{equ5.17}
E&=&\frac{1}{2}\mu\boldsymbol{v}^{2}-\frac{Gm\mu}{|\boldsymbol{x}|}-\frac{GQ^{1}Q^{2}}{3}\frac{\text{e}^{-m_{\text{s}}|\boldsymbol{x}|}}{|\boldsymbol{x}|}-G\sum_{l=2}^{\infty}\frac{(-1)^{l}}{l!}m^{1}\hat{M}^{2}_{I_{l}}\partial_{I_{l}}\left(\frac{1}{|\boldsymbol{x}|}\right)\notag\\
&&-\frac{G}{3}\sum_{l=1}^{\infty}\frac{(-1)^{l}}{l!}Q^{1}\hat{Q}^{2}_{I_{l}}\partial_{I_{l}}\left(\frac{\text{e}^{-m_{\text{s}}|\boldsymbol{x}|}}{|\boldsymbol{x}|}\right).
\end{eqnarray}
Just as in the situation of a planet orbiting the sun, the above two equations describe a spherical body moving around a massive central body. By using them within the framework of Newtonian $f(R)$
gravity, one can quantitatively analyze how the non-spherical mass distribution (e.g.,~oblateness) of the central body influences the motion of the spherical body. Therefore, Eqs.~(\ref{equ5.16}) and (\ref{equ5.17}) are applicable to analyzing numerous significant phenomena in the Solar system such as the effect of the solar multipole moment on the orbit of a planet.
\section{Summary and discussions~\label{Sec:sixth}}
In this paper, the inter-body dynamics of an isolated self-gravitating system comprising $N$ extended fluid bodies are studied in the Newtonian limit of $f(R)$ gravity. In order to make research results applicable to a wide range of many-body systems, we consider the system subject to two assumptions: 1) Each body is composed of a perfect fluid, and maintains approximate hydrostatic equilibrium; 2) No matter is being ejected or accreted by any body, and their typical sizes are far less than the typical separation between them. The two assumptions explicitly indicate that the inter-body and intra-body dynamics of the system are largely decoupled from each other so that the motions of the bodies are independent of the details of their internal states. This enables us to describe motions of the bodies using a set of coarse-grained variables characterizing each body as a whole, such as mass, center-of-mass position, spin angular momentum, and multipole moment. In this work, we successfully establish this coarse-grained description of the inter-body dynamics for the system.

The orbital dynamics of the system are first explored by performing a multipole analysis on the center-of-mass accelerations of the bodies, which presents a challenging and critical issue. As the first task of this paper, we  resolve the issue and derive the multipole expansion for the center-of-mass acceleration of each body in the system by applying the symmetric and trace-free formalism  in terms of irreducible Cartesian tensors~\cite{Thorne:1980ru,Blanchet:1985sp,Blanchet:1989ki,Damour:1990gj}. The expansion comprises two components: the Coulomb-type part and the Yukawa-type part. The former, identical to its counterpart in GR, is encoded by the products of the mass multipole moments of the body with those of other bodies, and in contrast, the latter, representing the modification to the Coulomb-type part introduced by $f(R)$ gravity, is encoded by the products of the scalar multipole moments of the body with those of other bodies. The multipole expansions for the center-of-mass accelerations of all the bodies form the foundation for understanding the orbital motion of the system within the framework of $f(R)$ gravity because they provide a complete set of equations of motion for the bodies once the mass and scalar multipole moments of the bodies are specified.

From the multipole expansion of each body's center-of-mass acceleration, several motion characteristics of the system can be identified. The monopole terms reveal that in the system composed of ideal spherical bodies, the presence of the scalar monopole term leads to the center-of-mass acceleration of a spherical body differing essentially from that of a point source. This is a key distinguishing feature of $f(R)$ gravity compared to GR (where spherical bodies coincide with point sources) and long-range massless scalar-tensor theories~\cite{Kopeikin2019} (where point-source equivalence depends subtly on the definitions of multipole moments). This fact underscores that in $f(R)$ gravity, it is the finite range of the massive scalar field that intrinsically governs the importance of body sizes in determining orbital motion. Furthermore, the higher-order terms indicate that in the system consisting of bodies exhibiting non-spherical distributions, each body's orbital motion is influenced by three factors: the deformation-induced gravitational potentials generated by other bodies, coupling between its own non-spherical mass distribution and the monopole potentials caused by other bodies, and interactions between its own higher-order multipole moments and those of other bodies. Given that the higher-order mass and scalar multipole moments of an object are often employed to characterize its deformation or non-spherical mass distribution, these three factors suggest that both the size and shape of the bodies within the system can have significant impacts on the orbital motion of each body under $f(R)$ gravity. Although these multipole couplings also exist in GR and massless scalar-tensor theories, in GR the mass dipole moment vanishes and higher-order mass multipole terms decay with order. In massless scalar-tensor theories, the scalar dipole moment is generally non-zero, but the scalar multipole terms still decay with increasing order. In contrast, within $f(R)$ gravity, the scalar dipole moment is non-zero, and more importantly, the scalar multipole terms do not decay with order because of the Yukawa kernel. Consequently, even if each body's size is far less than the typical separation between bodies, the equations of motion cannot be treated as a starting point of an approximation scheme in the same way as in GR.

To further explore the orbital dynamics of the system, its total momentum, energy, and angular momentum are next investigated. As the second task of this paper, we demonstrate that these quantities are conserved in $f(R)$ gravity at the Newtonian order, and also provide their expressions in the coarse-grained description of the system. The derivation of the expression for the total energy requires careful treatment because a multipole analysis of the total gravitational potential energy is necessary. To this end, we first derive the multipole expansion for the total gravitational potential energy of the system. It is shown that the interaction energies between different bodies arise from two types of couplings: one involving mass multipole moments and the other involving scalar multipole moments. Building on this finding, the expression for the total energy of the system is then obtained. The monopole-monopole potential energy reveals that the total energy of the system composed of ideal spheres is distinct from that of the system made up of point sources, which means that unlike in GR, the sizes of the spherical bodies make an important contribution to the system's total energy in $f(R)$ gravity. Apart from the monopole-monopole potential energy, the system's energy is also contributed by all the monopole-multipole and multipole-multipole potential energies between different bodies. These results explicitly indicate that the monopole-dipole and dipole-dipole potential energies are essential for the system's energy under $f(R)$ gravity, whereas they do not play a role in GR as the mass dipole moments of bodies are zero. These multipole couplings also appear in massless scalar-tensor theories. In those theories, scalar dipole moments are non-zero, but higher-order scalar multipole terms still decay with increasing order. In GR, by contrast, mass dipole moments vanish and higher-order mass terms decay. The distinctive feature of $f(R)$ gravity is that, due to the Yukawa kernel, scalar multipole terms do not decay with order, thereby violating the usual power-counting and truncation scheme.

The conservation of the system's total angular momentum implies that the spin angular momentum of an individual body is generally time-dependent. This makes it particularly meaningful to explore the spin dynamics of the system. As the third task of this paper, we derive the multipole expansion for time derivative of each body's spin angular momentum and provide the corresponding equation of motion governing its spin behavior in $f(R)$ gravity at the Newtonian order. The expansion also consists of the Coulomb-type and Yukawa-type parts that arise from the interactions between the multipole moments of the body and those of other bodies, which presents the similarity between spin and translational acceleration. Despite this resemblance, the distinctions between them can still be identified. Firstly, it is shown that the body's monopole moments have no effect on its spin's motion. Secondly, the definition of the scalar multipole moments involved in the expansion for the spin angular momentum needs a slight extension to incorporate the concept of weight  (cf.~Eq.~(\ref{equ4.53})) compared to the conventional definition employed in the expansion for the center-of-mass acceleration. These observations underscore the intrinsic nature of the spin angular momentum. With the multipole expansions for all bodies' spin angular momenta, we lay the groundwork for understanding spin motion of the system under $f(R)$ gravity.

Within the content of Newtonian $f(R)$ gravity, the three preceding tasks establish the core framework for the coarse-grained description of inter-body dynamics in an isolated self-gravitating system comprising $N$ extended fluid bodies. To apply the corresponding findings to studying the orbital dynamics of a two-body system, as the fourth task of this paper, we derive the effective one-body equation governing the relative motion between the two bodies and the expression for their total energy in $f(R)$ gravity at the Newtonian order. With these results,  a variety of significant astrophysical phenomena can be systematically addressed within $f(R)$ gravity models. For instance, as demonstrated in previous studies \cite{DeMartino:2018yqf, DeLaurentis:2018ahr}, the orbital motion of a test particle moving around a massive point body is investigated, and our results are able to extend the study to more general scenarios, specifically arbitrary two-body systems composed of point masses or spherical bodies, even when the two bodies' masses are comparable. In addition, our results are also applicable to studying how a massive central body's oblateness influences an orbiting sphere's trajectory via perturbative approach in celestial mechanics, which is a topic of interest in the many-body system. In light of this, our results could serve as a cornerstone for studying numerous important phenomena in the stellar systems within the context of $f(R)$ gravity.

As mentioned earlier, $f(R)$ gravity has emerged as a promising alternative to GR~\cite{Clifton:2011jh}. To systematically test $f(R)$ gravity, it is essential to develop the framework of celestial mechanics within this theoretical context. The results presented in this paper provide the foundational ingredients for constructing this framework in the Newtonian approximation of $f(R)$ gravity. In view of this, one of our subsequent tasks is to utilize the results of this paper to analyze various astrophysical phenomena so as to constrain the model parameters appearing in $f(R)$ gravity through observational data. Furthermore, the natural progression beyond the present Newtonian description is to formalize a self-consistent theory of relativistic celestial mechanics within $f(R)$ gravity. The methodologies introduced in this paper provide a direct pathway toward this goal, which is essential for establishing a complete post-Newtonian approximation. The successful construction of such an approximation will, in turn, enable significantly tighter and more systematic constraints on the parameters of $f(R)$ gravity. The effective field theory framework, as discussed in the Introduction, provides a natural tool to systematically incorporate the massive scalar mode at higher post-Newtonian orders in this construction. Moreover, while the present work assumes an isolated system embedded in a flat background spacetime, extending this framework to a more realistic cosmological FLRW background (as has been elegantly demonstrated in scalar-tensor cosmology, e.g., in Ref.~\cite{Galiautdinov2016}) remains a significant and interesting avenue for future research. Thus, the results and methodologies presented here constitute a critical foundation for the empirical testing of this alternative theory of gravity.


\acknowledgments{This work was supported by the National Natural Science Foundation of China (Grants Nos.~12105039, 12494574, and 12326602). This work was also supported by the Guangxi Talent Programs (``Young Seedling Talent'' and ``Highland of Innovation Talent'').}
\appendix\label{appendix}
\section{The weak-field and slow-motion approximation of $f(R)$ gravity~\label{Sec:appfirst}}
In Ref.~\cite{Wu:2021uws}, the metric for the external gravitational field of a spatially compact-supported source up to $1/c^3$ order in $f(R)$ gravity is provided in the form of multipole expansion.
To prepare for the derivations in Sec.~\ref{Sec:third}, we intend to revisit the WFSM approximation of $f(R)$ gravity and summarize the corresponding results in this Appendix.

In $f(R)$ gravity, motivated by the Landau-Lifshitz formulation of GR, the gravitational field amplitude $h^{\mu\nu}$ is defined by
\begin{eqnarray}
\label{equA1}h^{\mu\nu}:=\sqrt{-g}g^{\mu\nu}-\eta^{\mu\nu}
\end{eqnarray}
with $\eta^{\mu\nu}$ as the Minkowskian metric in a fictitious flat spacetime. Let $h^{\mu\nu}$ be a perturbation, and then, the linearized gravitational field equations of $f(R)$ gravity can be derived from Eqs.~(\ref{equ2.11}) and (\ref{equ2.12}),
\begin{eqnarray}
\label{equA2}H^{\mu\nu(1)}&=&R^{\mu\nu(1)}-\frac{1}{2}\eta^{\mu\nu}R^{(1)}+2a\eta^{\mu\nu}\square_{\eta}R^{(1)}-2a\partial^{\mu}\partial^{\nu}R^{(1)}=\kappa T^{\mu\nu},
\end{eqnarray}
where $\partial^{\rho}:=\eta^{\rho\sigma}\partial_{\sigma}$, $\square_{\eta}:=\eta^{\mu\nu}\partial_{\mu}\partial_{\nu}$, the superscript (1) represents the linear part of the corresponding quantity, and
$T^{\mu\nu}$ is the energy-momentum tensor of a source living in Minkowski spacetime.  The simplification of the linearized gravitational field equations relies on the gauge condition~\cite{Berry:2011pb,Naf:2011za}
\begin{eqnarray}
\label{equA3}\partial_{\mu}\tilde{h}^{\mu\nu}=0
\end{eqnarray}
with
\begin{eqnarray}
\label{equA4}\tilde{h}^{\mu\nu}:=h^{\mu\nu}+2a\eta^{\mu\nu}R^{(1)},
\end{eqnarray}
and by imposing it, the linearized gravitational field equations decouple into
\begin{eqnarray}
\label{equA5}&&\square_{\eta}\tilde{h}^{\mu\nu}=2\kappa T^{\mu\nu},\\
\label{equA6}&&\square_{\eta}R^{(1)}-m_{\text{s}}^{2}R^{(1)}=m_{\text{s}}^{2}\kappa T,
\end{eqnarray}
where
\begin{eqnarray}
\label{equA7}&&T:=\eta_{\mu\nu}T^{\mu\nu},\\
\label{equA8}&&m_{\text{s}}^2:=\frac{1}{6a}.
\end{eqnarray}
It can easily be seen that by means of the gauge condition~(\ref{equA3}),  the linearized gravitational field equations~(\ref{equA2}) are reduced to a manageable set of equations composed of Eqs.~(\ref{equA5}) and (\ref{equA6}).  With these results, if a physically meaningful solution to  Eqs.~(\ref{equA5}) and (\ref{equA6}) for a source is found, from Eqs.~(\ref{equA4}) and (\ref{equA8}), one can obtain the gravitational field amplitude $h^{\mu\nu}$ by using
\begin{eqnarray}
\label{equA9}h^{\mu\nu}&=&\tilde{h}^{\mu\nu}-\frac{1}{3m_{\text{s}}^2}\eta^{\mu\nu}R^{(1)},
\end{eqnarray}
and then, in the weak-field approximation, the metric for the gravitational field of the source will be provided by~\cite{Wu:2021uws}
\begin{eqnarray}
\label{equA10}g_{\mu\nu}=\eta_{\mu\nu}-\overline{h}_{\mu\nu},
\end{eqnarray}
where
\begin{eqnarray}
\label{equA11}\overline{h}_{\mu\nu}:=h_{\mu\nu}-\frac{1}{2}\eta_{\mu\nu}h,\quad h=\eta_{\mu\nu}h^{\mu\nu}.
\end{eqnarray}
Equation~(\ref{equA9}) shows that the metric is separated into two distinct components associated with the fields $\tilde{h}^{\mu\nu}$ and $R^{(1)}$, and to study their behaviors, the gauge transformations satisfied by them should be addressed. One can verify that under an infinitesimal coordinate transformation $x'^{\mu}= x^{\mu}+\varepsilon^{\mu}$, the gravitational field amplitude $h^{\mu\nu}$ and the linearized Ricci scalar $R^{(1)}$ satisfy
\begin{eqnarray}
\label{equA12}&&h'^{\mu\nu}=h^{\mu\nu}+\partial^{\mu}\varepsilon^{\nu}+\partial^{\nu}\varepsilon^{\mu}-\eta^{\mu\nu}\partial_{\alpha}\varepsilon^{\alpha},\\
\label{equA13}&&R'^{(1)}=R^{(1)},
\end{eqnarray}
and from them and Eq.~(\ref{equA4}), the transformation for the tensor field $\tilde{h}^{\mu\nu}$ is
\begin{eqnarray}
\label{equA14}&&\tilde{h}'^{\mu\nu}=\tilde{h}^{\mu\nu}+\partial^{\mu}\varepsilon^{\nu}+\partial^{\nu}\varepsilon^{\mu}-\eta^{\mu\nu}\partial_{\alpha}\varepsilon^{\alpha}.
\end{eqnarray}
The observation of Eqs.~(\ref{equA3}),  (\ref{equA5}), and (\ref{equA14}) reveals that the field $\tilde{h}^{\mu\nu}$  exhibits the behavior analogous to the gravitational field amplitude $h_{\text{GR}}^{\mu\nu}$ in GR, so it represents the massless propagation in $f(R)$ gravity. Thus, one could conclude that the introduction of the gauge condition~(\ref{equA3}) is primarily aimed at separating the massless tensor field $\tilde{h}^{\mu\nu}$ from the gravitational field amplitude $h^{\mu\nu}$. Regarding the scalar field $R^{(1)}$, its behavior is determined by the sign of parameter $m_{\text{s}}^2$ in Eq.~(\ref{equA6}). Since Eq.~(\ref{equA6}) reduces to the Klein-Gordon (KG) equation with external source when $m_{\text{s}}^2>0$, $R^{(1)}$ in this case should represent the massive propagation in $f(R)$ gravity. As indicated in Refs.~\cite{Capozziello:2007ms,DeMartino:2018yqf,DeLaurentis:2018ahr}, the scalar field $R^{(1)}$ in this scenario has significant applications in gravitational physics. In contrast, solutions to equation (\ref{equA6}) with a negative squared parameter might also be physically relevant, but they have not been extensively studied in the literature. In view of this, we will focus exclusively on the case of $m_{\text{s}}^2 > 0$ to ensure that our results have practical implications.

Next, in order to derive the gravitational potential in $f(R)$ gravity, Eqs.~(\ref{equA5}) and (\ref{equA6}) need to be further simplified by applying the slow-motion approximation. Given that the gravitational potential can be read off from the metric up to $1/c^2$ order, we assume that the metric outside a source takes the following form
\begin{equation}\label{equA15}
\left\{\begin{array}{ll}
\displaystyle g_{00}(t,\boldsymbol{x})&=\displaystyle -1+\frac{2}{c^{2}}\varPhi(t,\boldsymbol{x}),\smallskip\\
\displaystyle g_{0i}(t,\boldsymbol{x})&=\displaystyle 0,\smallskip\\
\displaystyle g_{ij}(t,\boldsymbol{x})&=\displaystyle \delta_{ij}+\frac{2}{c^{2}}\varPhi_{ij}(t,\boldsymbol{x}),
\end{array}\right.
\end{equation}
where $\varPhi(t,\boldsymbol{x})$ and $\varPhi_{ij}(t,\boldsymbol{x})$ are the scalar and tensor potentials, respectively. As implies in Refs.~\cite{Clifford2018,Eric2014}, $\varPhi(t,\boldsymbol{x})$ is just the gravitational potential in $f(R)$ gravity. Besides the metric,
it is shown from these references that under the slow-motion approximation, due to
\begin{equation}\label{equA16}
\frac{|\partial/\partial x^{0}|}{|\partial/\partial x^{i}|}\sim O\left(\frac{1}{c}\right),
\end{equation}
there is
\begin{eqnarray}
\label{equA17}\square_{\eta}=-\frac{\partial^{2}}{c^{2}\partial t^{2}}+\delta^{ij}\frac{\partial^{2}}{\partial x_{i}\partial x_{j}}\approx\delta^{ij}\frac{\partial^{2}}{\partial x_{i}\partial x_{j}}=\Delta
\end{eqnarray}
at the leading order, where $O(1/c^n)$ ($n$ is an integer) denotes the ``order of smallness'' of the corresponding quantity. Concerning the energy-momentum tensor, its components up to $1/c^2$ order fulfill~\cite{Stabile:2010zk,Damour:1990gj}
\begin{equation}
\label{equA18}\kappa T^{00}\sim O\left(\frac{1}{c^2}\right),\quad \kappa T^{0i}=0,\quad \kappa T^{ij}=0,
\end{equation}
from which, one could derive
\begin{equation}
\label{equA19}\kappa T=-\kappa T^{00}\sim O\left(\frac{1}{c^2}\right).
\end{equation}
These results, together with Eq.~(\ref{equA17}), suggest that up to $1/c^2$ order, Eqs.~(\ref{equA5}) and (\ref{equA6}) reduce to
a Poisson equation and a screened Poisson equation, namely,
\begin{eqnarray}
\label{equA20}&&\Delta\tilde{h}^{\mu\nu}=2\kappa T^{\mu\nu},\\
\label{equA21}&&\Delta R^{(1)}-m_{\text{s}}^2 R^{(1)}=m_{\text{s}}^2\kappa T.
\end{eqnarray}
These two equations serve as the starting point to derive the metric for a source in the WFSM approximation within $f(R)$ gravity, and they can be solved by applying the Green's function method.
The Green's functions of Eqs.~(\ref{equA20}) and (\ref{equA21}) are
\begin{eqnarray}
\label{equA22}&&\mathcal{G}_{\text{tensor}}(\boldsymbol{x};\boldsymbol{x}')=\frac{1}{4\pi|\boldsymbol{x}-\boldsymbol{x}'|},\\
\label{equA23}&&\mathcal{G}_{\text{scalar}}(\boldsymbol{x};\boldsymbol{x}')=\frac{\text{e}^{-m_{\text{s}}|\boldsymbol{x}-\boldsymbol{x}'|}}{4\pi|\boldsymbol{x}-\boldsymbol{x}'|},
\end{eqnarray}
and they satisfy~\cite{Wu:2017huang,Wu:2022akq}
\begin{eqnarray}
\label{equA24}&&\nabla^2 \mathcal{G}_{\text{tensor}}(\boldsymbol{x};\boldsymbol{x}')=-\delta^{3}(\boldsymbol{x}-\boldsymbol{x}'),\\
\label{equA25}&&\left(\nabla^2-m_{\text{s}}^{2}\right)\mathcal{G}_{\text{scalar}}(\boldsymbol{x};\boldsymbol{x}')=-\delta^{3}(\boldsymbol{x}-\boldsymbol{x}').
\end{eqnarray}
Based on the above four equations, the solutions to Eqs.~(\ref{equA20}) and (\ref{equA21}) can be written as
\begin{eqnarray}
\label{equA26}&&\tilde{h}^{\mu\nu}(t,\boldsymbol{x})=\int\mathcal{G}_{\text{tensor}}(\boldsymbol{x};\boldsymbol{x}')\left(-2\kappa T^{\mu\nu}(t,\boldsymbol{x}')\right)\text{d}^{3}x',\\
\label{equA27}&&R^{(1)}(t,\boldsymbol{x})=\int\mathcal{G}_{\text{scalar}}(\boldsymbol{x};\boldsymbol{x}')\left(-m_{\text{s}}^2\kappa T(t,\boldsymbol{x}')\right)\text{d}^{3}x',
\end{eqnarray}
and by further employing Eqs.~(\ref{equA18}) and (\ref{equA19}), they can be rewritten as
\begin{equation}\label{equA28}
\left\{\begin{array}{ll}
\displaystyle \tilde{h}^{00}(t,\boldsymbol{x})&=\displaystyle -\frac{4G}{c^2}\int\frac{T^{00}(t,\boldsymbol{x}')}{c^2|\boldsymbol{x}-\boldsymbol{x}'|}
\text{d}^{3}x',\smallskip\\
\displaystyle \tilde{h}^{0i}(t,\boldsymbol{x})&=\displaystyle0,\smallskip\\
\displaystyle \tilde{h}^{ij}(t,\boldsymbol{x})&=\displaystyle0
\end{array}\right.
\end{equation}
and
\begin{eqnarray}
\label{equA29}&& R^{(1)}(t,\boldsymbol{x})=\frac{2Gm_{\text{s}}^{2}}{c^2}\int\frac{T^{00}(t,\boldsymbol{x}')\text{e}^{-m_{\text{s}}|\boldsymbol{x}-\boldsymbol{x}'|}}{c^2|\boldsymbol{x}-\boldsymbol{x}'|}\text{d}^{3}x'.
\end{eqnarray}
To present the metric in a neat form, the Coulomb-like and Yukawa-like  potentials are defined,
\begin{eqnarray}
\label{equA30}&&\displaystyle U(t,\boldsymbol{x})=\displaystyle G\int\frac{T^{00}(t,\boldsymbol{x}')}{c^2|\boldsymbol{x}-\boldsymbol{x}'|}\text{d}^{3}x',\\
\label{equA31}&&\displaystyle Y(t,\boldsymbol{x})=\displaystyle \frac{G}{3}\int\frac{T^{00}(t,\boldsymbol{x}')\text{e}^{-m_{\text{s}}|\boldsymbol{x}-\boldsymbol{x}'|}}{c^2|\boldsymbol{x}-\boldsymbol{x}'|}\text{d}^{3}x'
\end{eqnarray}
and then, there are
\begin{eqnarray}
\label{equA32}&&\displaystyle\tilde{h}^{00}(t,\boldsymbol{x})=-\frac{4U(t,\boldsymbol{x})}{c^2},\\
\label{equA33}&&\displaystyle R^{(1)}(t,\boldsymbol{x})=\frac{6m_{\text{s}}^{2}Y(t,\boldsymbol{x})}{c^2}.
\end{eqnarray}
Thus, with them, from Eq.~(\ref{equA9}), the gravitational field amplitude $h^{\mu\nu}$ and its trace are expressed as
\begin{equation}\label{equA34}
\left\{\begin{array}{ll}
\displaystyle h^{00}(t,\boldsymbol{x})&=\displaystyle -\frac{4U(t,\boldsymbol{x})}{c^2}+\frac{2Y(t,\boldsymbol{x})}{c^2},\smallskip\\
\displaystyle h^{0i}(t,\boldsymbol{x})&=\displaystyle0,\smallskip\\
\displaystyle h^{ij}(t,\boldsymbol{x})&=\displaystyle-\delta^{ij}\frac{2Y(t,\boldsymbol{x})}{c^2}
\end{array}\right.
\end{equation}
and
\begin{eqnarray}
\label{equA35}h(t,\boldsymbol{x})&=&\frac{4U(t,\boldsymbol{x})}{c^2}-\frac{8Y(t,\boldsymbol{x})}{c^2}.
\end{eqnarray}
After substituting the above results in Eqs.~(\ref{equA10}) and (\ref{equA11}), we finally achieve the metric, presented in the form of Eq.~(\ref{equA15}), for the external gravitational field of the source, where the scalar and tensor potentials $\varPhi(t,\boldsymbol{x})$ and $\varPhi_{ij}(t,\boldsymbol{x})$ are, respectively,
\begin{eqnarray}
\label{equA36}&&\varPhi(t,\boldsymbol{x})=U(t,\boldsymbol{x})+Y(t,\boldsymbol{x}),\\
\label{equA37}&&\varPhi_{ij}(t,\boldsymbol{x})=\delta_{ij}\left[U(t,\boldsymbol{x})-Y(t,\boldsymbol{x})\right].
\end{eqnarray}
The above two equations explicitly demonstrate that the temporal and spatial components of the metric are governed by two distinct scalar potentials, namely $U+Y$ and $U-Y$, respectively. This unequal treatment of spatial and temporal metric perturbations is a well-known hallmark of the weak-field limit in metric $f(R)$ and scalar-tensor gravity, as has been extensively discussed in the literature for general or point-like sources~\cite{Capozziello2010,Martino2018,Toniato2018}. Our results explicitly realize and generalize this classical feature for extended bodies.
Obviously, when $f(R)=R$, Eqs.~(\ref{equ2.12}) and (\ref{equA8}) imply that $m_{\text{s}}\rightarrow+\infty$, which results in that the Yukawa-like potential $Y(t,\boldsymbol{x})$ goes to zero. Consequently, the scalar and tensor potentials $\varPhi(t,\boldsymbol{x})$ and $\varPhi_{ij}(t,\boldsymbol{x})$ reduce to
\begin{eqnarray}
\label{equA39}&&\varPhi^{\text{GR}}(t,\boldsymbol{x})= U(t,\boldsymbol{x}),\\
\label{equA40}&&\varPhi^{\text{GR}}_{ij}(t,\boldsymbol{x})=\delta_{ij}U(t,\boldsymbol{x}),
\end{eqnarray}
and the corresponding metric in this case recovers the one in GR. Equations~(\ref{equA15})--(\ref{equA37}) constitute the most fundamental ingredient of the WFSM approximation framework for $f(R)$ gravity, and within this framework, people can readily obtain the metric for the gravitational field of a spatially compact-supported source up to $1/c^{2}$ order.

\end{document}